\documentclass[twocolumn]{aastex631}

\usepackage{graphicx}	
\usepackage{amsmath}	
\usepackage{amssymb}	
\usepackage{ulem}

\shorttitle{Ram-Pressure Stripped Galaxies in A2744 and A370}
\shortauthors{C. Bellhouse et al.}

% ...

\begin{document}

\title{Locations and Morphologies of Jellyfish Galaxies in A2744 and A370}

\email{callum.bellhouse@inaf.it}
\author[0000-0002-6179-8007]{Callum Bellhouse}
\affiliation{INAF-Padova Astronomical Observatory,\\ Vicolo dell’Osservatorio 5, I-35122 Padova, Italy}

\author[0000-0001-8751-8360]{Bianca Poggianti}
\affiliation{INAF-Padova Astronomical Observatory,\\ Vicolo dell’Osservatorio 5, I-35122 Padova, Italy}

\author[0000-0002-1688-482X]{Alessia Moretti}
\affiliation{INAF-Padova Astronomical Observatory,\\ Vicolo dell’Osservatorio 5, I-35122 Padova, Italy}

\author[0000-0003-0980-1499]{Benedetta Vulcani}
\affiliation{INAF-Padova Astronomical Observatory,\\ Vicolo dell’Osservatorio 5, I-35122 Padova, Italy}

\author[0000-0002-4382-8081]{Ariel Werle}
\affiliation{INAF-Padova Astronomical Observatory,\\ Vicolo dell’Osservatorio 5, I-35122 Padova, Italy}

\author[0000-0002-7296-9780]{Marco Gullieuszik}
\affiliation{INAF-Padova Astronomical Observatory,\\ Vicolo dell’Osservatorio 5, I-35122 Padova, Italy}

\author[0000-0002-3585-866X]{Mario Radovich}
\affiliation{INAF-Padova Astronomical Observatory,\\ Vicolo dell’Osservatorio 5, I-35122 Padova, Italy}

\author[0000-0003-2150-1130]{Yara Jaff\'e}
\affiliation{Instituto de F\'isica y Astronom\'ia, Universidad de Valpara\'iso, Avda. Gran Breta\~na 1111 Valpara\'iso, Chile}

\author[0000-0002-7042-1965]{Jacopo Fritz}
\affiliation{Instituto de Radioastronomia y Astrofisica, UNAM, Campus Morelia, AP 3-72, CP 58089, Mexico}

\author[0000-0003-1581-0092]{Alessandro Ignesti}
\affiliation{INAF-Padova Astronomical Observatory,\\ Vicolo dell’Osservatorio 5, I-35122 Padova, Italy}

\author[0000-0002-8372-3428]{Cecilia Bacchini}
\affiliation{INAF-Padova Astronomical Observatory,\\ Vicolo dell’Osservatorio 5, I-35122 Padova, Italy}

\author[0000-0002-8238-9210]{Neven Tomi\v{c}i\'c}
\affiliation{INAF-Padova Astronomical Observatory,\\ Vicolo dell’Osservatorio 5, I-35122 Padova, Italy}

\author[0000-0001-5492-1049]{Johan Richard}
\affiliation{Univ Lyon, Univ Lyon1, Ens de Lyon, CNRS, Centre de Recherche Astrophysique de Lyon UMR5574, F-69230, Saint-Genis-Laval, France}

\author[0000-0001-9976-1237]{Genevi\`eve Soucail}
\affiliation{Institut de Recherche en Astrophysique et Plan\'etologie (IRAP), Universit\'e de Toulouse, CNRS, UPS, CNES,14 Av. Edouard Belin, 31400 Toulouse, France}


\begin{abstract}

We present a study of the orbits, environments and morphologies of 13 ram-pressure stripped galaxies in the massive, intermediate redshift (z$\sim0.3-0.4$) galaxy clusters A2744 and A370, using MUSE integral-field spectroscopy and HST imaging from the Frontier Fields Program. We compare different measures of the locations and morphologies of the stripped sample with a sample of 6 poststarburst galaxies identified within the same clusters, as well as the general cluster population. We calculate the phase space locations of all cluster galaxies and carry out a substructure analysis, finding that the ram-pressure stripped galaxies in A370 are not associated with any substructures, but are likely isolated infalling galaxies. In contrast, the ram-pressure stripped galaxies in A2744 are strictly located within a high-velocity substructure, moving through a region of dense X-ray emitting gas. We conclude that their ram-pressure interactions are likely to be the direct result of the merger between two components of the cluster.
Finally, we study the morphologies of the stripped and poststarburst galaxies, using numerical measures to quantify the level of visual disturbances. We explore any morphological deviations of these galaxies from the cluster population, particularly the weaker cases which have been confirmed via the presence of ionised gas tails to be undergoing ram-pressure stripping, but are not strongly visually disturbed in the broad-band data. We find that the stripped sample galaxies are generally divergent from the general cluster sample, with poststarburst galaxies being intermediary in morphology between stripped galaxies and red passive cluster members.

\end{abstract}




\section{Introduction}

The study of galaxy interactions within clusters is critical to understanding the growth and development of galaxies. Many different processes can affect or disrupt a galaxy's gas content, which can have profound effects on its subsequent evolution.

Since the early work of \citet{Butcher1978a,Butcher1978b}, it has been understood that the color distribution of a galaxy population evolves with redshift, with greater fractions of blue galaxies at higher redshifts in comparison to the local universe. Many works since then have proposed different mechanisms, both internal to a galaxy, or resulting from interactions with its environment, that can act to transform galaxies and may contribute to the quenching of the global population of galaxies throughout cosmic time.

The environmental processes that act upon cluster galaxies can be divided into gravitational and hydrodynamical effects. The former include \textit{tidal interactions} \citep{Spitzer1951,Toomre1977,Tinsley1979,1983ApJ...264...24M,Mihos1994,Springel2000}, caused by direct gravitational interaction between galaxies, and \textit{harassment} \citep{1996Natur.379..613M,1998ApJ...495..139M}, resulting from the cumulative effect of many high-speed close approach encounters between cluster members. In both gravitational processes, the stellar \textit{and} gas components of the galaxies are affected.

On the other side of the coin lie the processes known as hydrodynamical interactions, which primarily impact the gas component with little to no impact on the prior stellar population. Such effects include the removal of the outer gas reservoir of the affected galaxy via \textit{starvation/strangulation} \citep{Larson1980,Balogh2000}, and in more extreme cases, the stripping of the internal gas component from the galaxy in the process known as \textit{ram-pressure stripping} (hereafter RPS).

First discussed in \citet{1972ApJ...176....1G}, RPS is one of the most efficient mechanisms \citep{Boselli2006} which can abruptly quench star formation and greatly disrupt a galaxy's gas content. RPS can occur when a galaxy moves sufficiently quickly through the dense intracluster medium (hereafter ICM) of galaxy clusters. The ram-pressure effect scales with environmental density and galaxy velocity, preferentially affecting galaxies on steep, radially infalling orbits \citep{Jaffe2018}. A galaxy's ability to retain its gas is dependent on its stellar and gas mass surface densities as well as the mass of its dark matter halo, with more massive galaxies able to weather the effect and retain some gas for longer than their less massive counterparts. The ram-pressure acts upon the gas component of a galaxy with only subtle, indirect influence on the existing stellar component \citep{Smith2012}. Observational effects can include the formation of tails of stripped gas trailing behind the galaxy \citep{Fumagalli2014}, compression of the leading edge of the disc by the ICM \citep{2001ApJ...561..708V}, and the onset of star formation in the tails with the condensation of star-forming clumps \citep{Kenney2014}. In addition, it has recently been shown \citep{Bellhouse2021} that unwinding of the galaxy's spiral arms can occur during the early stages of RPS, due to removal of material from the outside edges of the disc.

Processes which induce abrupt quenching in cluster galaxies give rise to the population of so-called post starburst (hereafter PSB) galaxies. PSB galaxies exhibit low to negligible star formation characterised by a lack of nebular emission lines, whilst exhibiting strong Balmer absorption lines indicative of relevant star formation in the past $\sim$ 1 Gyr. Previous studies have pointed towards RPS being responsible, at least in part, for the formation of PSB galaxies in clusters. Evidence includes the correlation of their locations with substructures in the ICM \citep{Poggianti2004}, the similarity in the spectral features of recently quenched regions within RPS galaxies compared with PSB galaxies \citep{Gullieuszik2017, Werle2022}, and evidence of very recent stripping in certain PSB galaxies \citep{Werle2022}.

Although RPS is known to be one of the most efficient mechanisms influencing the evolution of cluster populations, its relative contribution within the context of galaxy evolution in the universe is only recently starting to be quantified \citep{Vulcani2022}. Many examples of RPS galaxies have been presented to date, including in large surveys of visually selected RPS candidates such as \citet{Poggianti2016} and \citet{Roberts2020} at low redshifts, \citet{McPartland2016} from $0.3<z<0.7$ and as part of wider surveys such as the Grism Lens-Amplified Survey from Space \citep[GLASS,][]{Treu2015b,Vulcani2016, Vulcani2017}. This has given us an overview of the stripping processes throughout the recent history of the universe and beyond, and newer studies have continued to probe outside the local universe \citep{boselli2019b,Kalita2019,Stroe2020,Durret2021,Ebeling2014}. Extending the known sample to higher redshifts opens up the opportunity to better understand the impact of stripping on the evolution of clusters, and an insight into the influence of stripping throughout the history of clusters in the universe today.

A powerful tool in diagnosing and understanding the processes which occur within and around a galaxy is Integral Field Unit (IFU) spectroscopy,
which enables the exploration of both the spatial and spectral information of the observed objects.
This allows properties such as star formation rates and star formation histories to be traced across a galaxy, and ionized gas to be mapped in location and velocity throughout the galaxy and its tails, making it an extremely useful instrument when characterizing RPS interactions. IFU observations have proven extremely useful to studies of RPS \citep{Merluzzi2013,Fumagalli2014,2016MNRAS.455.2028F,Poggianti2017}, probing the kinematics and ongoing processes within galaxies during infall and leading to new discoveries about the resulting effects of ram-pressure interactions on galaxies \citep{Poggianti2017b,Bellhouse2021}. In this analysis we will utilize observations gathered by the MUSE (Multi Unit Spectroscopic Explorer, \citealt{2010SPIE.7735E..08B}) IFU at the European Southern Observatory (ESO) Very Large Telescope.






In this work, we focus on two clusters, Abell 2744 and Abell 370, located around $0.3<z<0.4$, which are both post-mergers, and which are the first two clusters to have been observed within the MUSE Guaranteed Time Observations program.

The aim of this study is to investigate the distribution of identified RPS galaxies in A2744 and A370, both in phase space and within the context of the cluster substructures. We also aim to compare different morphological parameters to test whether these confirmed RPS galaxies could be detectable using automated methods, which could be applied to other frontier fields clusters in a future study.

This paper is part of a series of works which aim to characterise the process of stripping within frontier fields clusters, including \citet{Moretti2022} and \citet{Werle2022}.



The paper is structured as follows. Section \ref{sec:data} describes the data used in this analysis as well as the sample selection process. In section \ref{sec:phasespace+ds} we outline the phase space and substructure detection techniques used to analyse the distributions of galaxies in the clusters. In section \ref{sec:xray+lensing}, we compare the locations of the RPS and PSB galaxies with X-ray and mass maps of the clusters. Section \ref{sec:morphology} describes an analysis of the galaxy morphologies, comparing the RPS and PSB galaxies with the sample of red and blue cluster galaxies, using an array of different morphology measures. Finally, in section \ref{sec:discussion} we summarise and interpret the results of the work.

\section{Observations and Data}\label{sec:data}

\subsection{Clusters}\label{sec:clusters}

Abell 2744 (z=0.3064, $\sigma=1497 \mathrm{km\,s}^{-1},$ \citealt{Owers2011}, hereafter A2744) is a merging cluster with a virial mass of $7.4\times10^{15}\mathrm{M}_\odot$, mostly comprised of two distinct components with v$_{\rm pec} = -1308 \pm 161 \mathrm{km\,s}^{-1}$, $\sigma =1277\pm189\mathrm{km\,s}^{-1}$ and v$_{\rm pec} =2644\pm72\mathrm{km\,s}^{-1}$, $\sigma = 695 \pm 76 \mathrm{km\,s}^{-1}$ \citep{Mahler2018}. A2744 is in a particularly dynamical state due to its merging history, with a significant blue galaxy excess of $2.2\pm0.3$ times that of nearby clusters in the same core regions \citep{Owers2011}. The cluster's merging history is of particular interest, as it provides a valuable insight into the link between cluster mergers and galaxy star formation activity. An increased fraction of starbursting blue galaxies driven by interactions with the disturbed ICM is a keen candidate for a contributor to the scatter in the Butcher-Oemler effect \citep{Kauffmann1995,Miller2006}. The complex merging history of A2744 is explored in detail using X-ray and optical spectroscopy in \citet{Owers2011}, who identify two major substructures, the Northern Core (NC), and the Southern Minor Remnant Core (SMRC) within the cluster, as well as a region labelled the Central Tidal Debris (CTD), which is close in projection to the SMRC but exhibits a velocity close to that of the NC and propose a scenario of a post-core-passage major merger in addition to an interloping minor merger, with the CTD being a region stripped from the NC by the interaction. The locations of each of these regions are shown in Figure~\ref{fig:A2744_footprint} in Appendix~\ref{sec:appendix}, for context.

Abell 370 (z=0.375, $\sigma=1789\mathrm{km\,s}^{-1}$ \citealt{Richard2021}, hereafter A370) is a historically significant cluster both within the context of galaxy evolution studies \citep{1984ApJ...285..426B,Dressler1997,Dressler1999}, and also for the study of gravitational lensing, as it contains one of the first observations of a giant-arc lens system \citep{Lynds1986,Soucail1987,Soucail1988}. The cluster has a total virial mass of $\mathrm{M}_{\rm vir}$ = $3.3 \times 10^{15} \mathrm{M}_\odot$ from weak lensing measurements \citep{Umetsu2011a,Umetsu2011b}. The cluster exhibits a bimodal distribution of galaxies consistent with a major merger \citep{Richard2010} of two progenitor clusters with masses $\mathrm{M}_{\rm vir}$ = $1.7 \times 10^{15} \mathrm{M}_\odot$ and $\mathrm{M}_{\rm vir}$ = $1.6 \times 10^{15} \mathrm{M}_\odot$ \citep{Molnar2020}. The centres of the X-ray and dark matter in both peaks are fairly close in comparison to similar merging clusters which 
suggests that the merger axis is predominantly along the line of sight \citep{Richard2010}.
The northern and southern brightest cluster galaxies (BCGs) are located at z=0.3780 and z=0.3733 respectively \citep{Lagattuta2019}, indicating a separation of 1024$\mathrm{km\,s^{-1}}$ \citep{Molnar2020}. Unlike A2744, the distribution of velocities of the cluster members do not exhibit a distinctly bimodal distribution, suggesting that the merging clusters may have already experienced a previous passage leading to mixing of the populations \citep{Lagattuta2019}.


\subsection{Data}

In this study, we utilize data from the MUSE Lensing Cluster GTO program \citep{Bacon2017,Richard2021}. The observations cover the central regions of clusters with single pointings or mosaics, with effective exposure times from 2 hours up to 15 hours under very good seeing conditions ($\sim0''.6$). The full set of clusters observed with MUSE, described in \citet{Richard2021}, is compiled from the MAssive Clusters Survey \citep[MACS,][]{Ebeling2001}, Frontier Fields \citep[FF,][]{Lotz2017}, GLASS \citep[][]{Treu2015b} and the Cluster Lensing and Supernova survey with Hubble \citep[CLASH,][]{Postman2012} programs. In the case of A2744, the MUSE data consist of a $2\times2$ mosaic of GTO observations, with a field of view (FoV) of $\sim2'\times2'$ ($2' = 0.27\mathrm{R}_{200}$) centered on RA= $0\mathrm{h}~14'~20.952''$ and DEC = $-30^\circ~23'~53.88''$ covering the region that includes the southern and central structures but excludes the northern core and interloper (The MUSE FoV of A2744 is shown overlaid on the HST F606W image in Figure~\ref{fig:A2744_footprint} in Appendix~\ref{sec:appendix}). For A370, the data consist of a $2\times2$ mosaic of observations centered on RA= $2\mathrm{h}~39'~53.111''$ DEC=-$1^\circ~34'~55.77''$, which is an expansion on the single pointing of the GTO program, extending the coverage to a $\sim2'\times2'$ ($2' = 0.24\mathrm{R}_{200}$) region of the cluster \citep{Lagattuta2019} (The MUSE FoV of A370 is shown overlaid on the HST F606W image in Figure~\ref{fig:A370_footprint} in Appendix~\ref{sec:appendix}). The complete data analysis of the cluster sample and the redshift catalogs are given in \citet{Richard2021}. The HST data are comprised of WFPC2, ACS/WFC and WFC3-IR images which cover the MUSE observations, sourced from High-Level Science Product (HLSP) images from the CLASH and FF repositories. In this analysis we use only the data from HST which overlap the MUSE observations in the case of both clusters. In this particular analysis, we make use of the F435W, F606W and F814W observations from the FF data, full details of the observations are given in the FF survey paper \citep{Lotz2017}.

For comparison with the cluster mass distribution, we also make use of the Clusters As TelescopeS \citep[CATS,][]{Jullo2009,Richard2014,Jauzac2015a,Jauzac2015b,Limousin2016,Lagattuta2017,Mahler2018} mass surface density model. This model is produced using the \textsc{lenstool} code which uses the positions, magnitudes, shapes, multiplicity and redshifts of lensed objects to derive the mass distribution of the cluster. The overall mass distribution is calculated as a superposition of the smooth, global cluster potential and smaller individual substructures associated with bright cluster member galaxies. The full methodology of the technique is presented in \citet{Jullo2009}.

We also utilize X-ray images based on Chandra data, described in \cite{Mantz2010} \citep[see also][]{vonderLinden2014,Vulcani2017} in order to compare the cluster gas distribution to the locations of the observed RPS and PSB galaxies. The X-ray images use the 0.7-2.0~keV energy band observations, which is the preferred range for tracing gas mass, as the emissivity in this range is largely insensitive to the gas temperature \citep{Mantz2010}.

\subsection{Cluster Membership and Galaxy Colours}\label{sec:membership}

In order to have an overview of the cluster population against which we can later compare RPS and PSB galaxies, we identify the sample of cluster galaxies within the observed central region by their velocities, and highlight color-magnitude selected red and blue galaxies in order to contextualize our samples within the different cluster populations.

We first extracted the sample of confirmed cluster members by calculating the peculiar velocity of each galaxy using the \citet{Richard2021} spectroscopic redshifts, and selecting galaxies within a specified threshold of the cluster velocity, which was set at $\pm 3\times(1+\mathrm{z})\sigma$, where $\sigma$ denotes the cluster velocity dispersion.. 

\begin{figure*}
    \centering
    \includegraphics[width=0.49\textwidth]{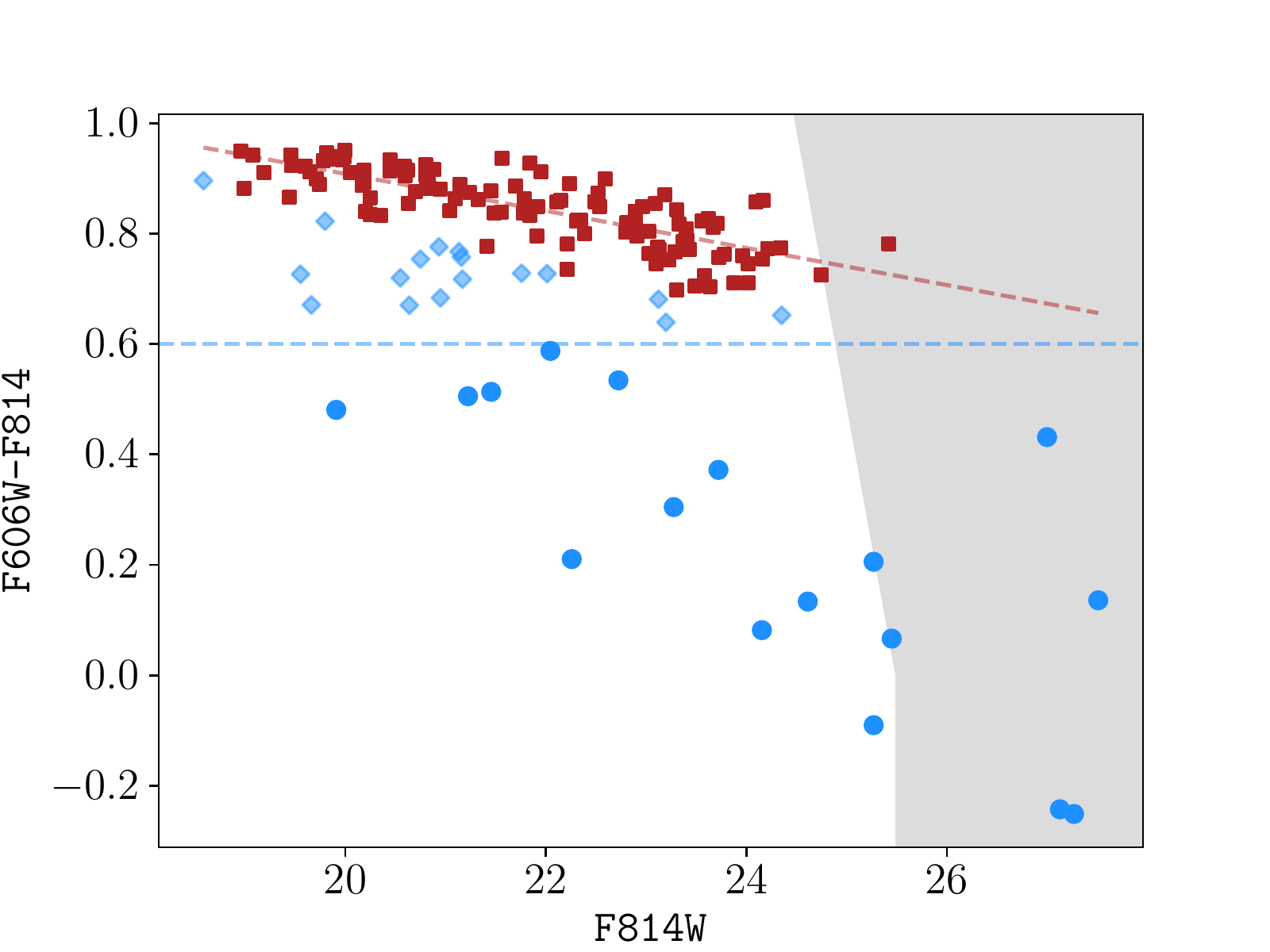}
    \includegraphics[width=0.49\textwidth]{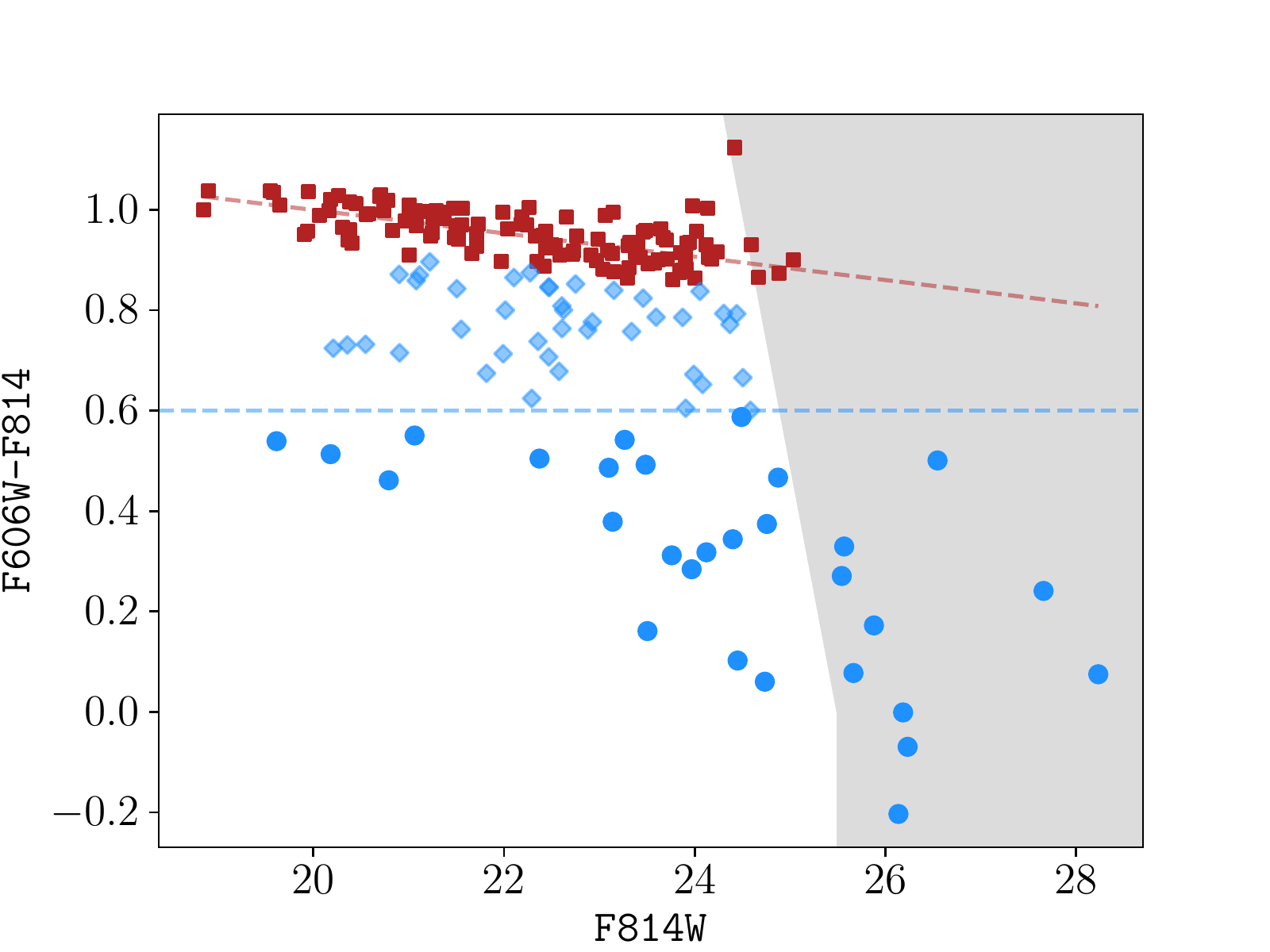}
    \caption{Color-Magnitude diagrams for A2744 (\textit{left}) and A370 (\textit{right}), showing all cluster members within the central regions of each cluster covered by the observations. Red galaxies are classified as described in the text (Section~\ref{sec:membership}) and marked as red squares. Other galaxies are divided into those with F606W-F814W$>0.6$, marked as light blue diamonds, and those with F606W-F814W$<0.6$, marked as blue circles. The red dashed line denotes the linear regression fitted to the red sequence galaxies. The grey shaded region indicates the magnitude limit of 25.5 applied in both the F606W and F814W bands.}
    \label{fig:CM_both}
\end{figure*}

For both A2744 and A370, we subdivided the velocity-cut sample into red and blue galaxies using their distribution in F606W-F814W vs F814W color-magnitude space, shown in Figure~\ref{fig:CM_both}.
For each cluster individually, we employed Gaussian mixture models (GMM) to extract two groupings of objects in the color-magnitude space, which closely corresponded to the red sequence and blue cloud. The Gaussian mixture model yields a probability with which each object belongs to either group. We fitted the red sequence by selecting galaxies with a probability $>0.9$ of belonging to the upper group of the color-magnitude space, and fit a linear regression line to this sample, marked by the red dashed line in Figure~\ref{fig:CM_both}.

We then assigned to the red sample all galaxies above the fitted red sequence line, as well as any galaxies below the line with a $>0.9$ probability of belonging to the upper group extracted from GMM. All remaining galaxies in the color-magnitude space were then assigned to the blue sample. We further subdivided the blue galaxies into those with F606W-F814W$<0.6$, marked as blue in the figure, and those with intermediate colors, marked in light blue.

For A2744, we spectroscopically confirm 158 cluster members. Of the 148 of these which have good magnitude measurements in F606W and F814W, 114 galaxies are on the red sequence.
For A370, we confirm 248 cluster members, 186 with good magnitude measurements in F606W and F814W, of which 116 are on the red sequence.

In the full sample across both clusters, the faintest red sequence galaxy we detect has an F814W magnitude of 25.4, and the faintest blue galaxy we detect has an F814W magnitude of 28.2. For the rest of the analysis, we applied a magnitude limit of 25.5 to both clusters in both F814W and F606W filters, which is close to our detection limit for red galaxies, but does not exclude any of our RPS or PSB sample. The magnitude limit is shown by the grey shaded region in both panels of Figure \ref{fig:CM_both}.


\subsection{Visual identification and sample selection}\label{sec:ident}

We use the samples of galaxies identified in \citet{Moretti2022}, who selected RPS and PSB galaxies based on visual inspection of the optical HST data and MUSE spectra simultaneously. Potential RPS galaxies were selected based on the presence of unilateral tails/debris exhibiting emission lines in the MUSE data. Some of the selected galaxies also exhibited tails in the HST data, which were confirmed to be associated with the galaxy from the MUSE redshifts. 
The PSB galaxies were selected based on their spectral features, targeting objects lacking emission lines associated with ongoing star-formation but exhibiting strong Balmer lines in absorption. 
This classification can be biased against objects that have some ionized gas due to processes other than star-formation, however, our spatially resolved data allows us to identify these cases.
This is the case of A2744\_07, where we find centrally concentrated emission associated with an AGN \citep[see][for details]{Werle2022}.
The full details of the RPS and PSB sample selection processes are outlined in \citet{Moretti2022}. A focused analysis of the PSB galaxies in these clusters, alongside others, is presented in \citet{Werle2022}. 

From \citet{Moretti2022}, there are 6 RPS galaxies as well as 4 PSB galaxies within the MUSE field of view of A2744. We note, however, two galaxies of interest, which show features of being both RPS and PSB \citep{Werle2022}. One of the RPS sample galaxies, A2744\_01, has a clear tail in $\mathrm{H}\alpha$ but no emission lines within the disk, suggesting that it is in an intermediate phase between the two types. In addition, one of the PSB galaxies, A2744\_07, has traces of extraplanar emission, suggestive of a tail from a past stripping event. We primarily classify A2744\_01 and A2744\_07 as RPS and PSB respectively, but highlight these galaxies to distinguish them in the rest of the analysis. The locations of the selected galaxies in A2744 are shown marked on the cluster in Figure~\ref{fig:A2744_footprint} in Appendix~\ref{sec:appendix}.

In the case of A370, 7 galaxies were visually identified with signs of RPS, alongside 2 PSB galaxies. Two of the galaxies are also noted in other MACS and Frontier Fields works, A370\_01 is highlighted as an extreme case of RPS in \citet{Ebeling2019}, and A370\_08 is noted in \citet{Lagattuta2019} as ID 8006 along with an associated clump of stripped material, ID CL49. The locations of the selected galaxies in A370 are shown marked on the cluster in Figure~\ref{fig:A370_footprint} in Appendix~\ref{sec:appendix}.

\section{Phase space and substructure analysis}\label{sec:phasespace+ds}

In order to explore the cluster environments and to understand the nature of the stripping process for each of the galaxies in our sample, we build a picture of the structure of each cluster and the orbits of the galaxies within using phase space maps and Dressler-Shectman \citep[][hereafter DS]{1988AJ.....95..985D} tests. In general, these diagnostics allow us to better understand how galaxies are interacting with the host cluster, such as the types of orbits they are on and whether they are associated with a group or substructure. Since both clusters are undergoing merging activity, it is particularly important to understand how these galaxies are situated within their environments.


\subsection{Phase Space Analysis}\label{sec:phasespace}

The nature of a galaxy's orbit gives us a useful measure of the likelihood with which it will experience RPS, which is far more common in galaxies passing close to the cluster center at high velocity.

In order to investigate the orbits of the galaxies in our sample, we produced plots of their locations in cluster-projected position velocity phase space \citep{HernandezFernandez2014}. These diagnostics reveal the nature of the galaxies' orbits, allowing us to determine whether they lie on more circular orbits or steep, plunging radial orbits, conducive to RPS \citep{Jaffe2015}. Typically, galaxies at low projected clustercentric radii with high line-of-sight velocities are likely to be within this regime of orbits \citep{Jaffe2018}, infalling steeply into their host cluster and experiencing RPS.

For the center of A370, we follow the definition given by \citet{Lah2009}, who use the mid-point between the northern and southern BCGs. We also use the R$_{200}$ radius of 2.57 Mpc from \citet{Lah2009}.

\begin{figure*}
    \centering
    \includegraphics[width=0.49\textwidth]{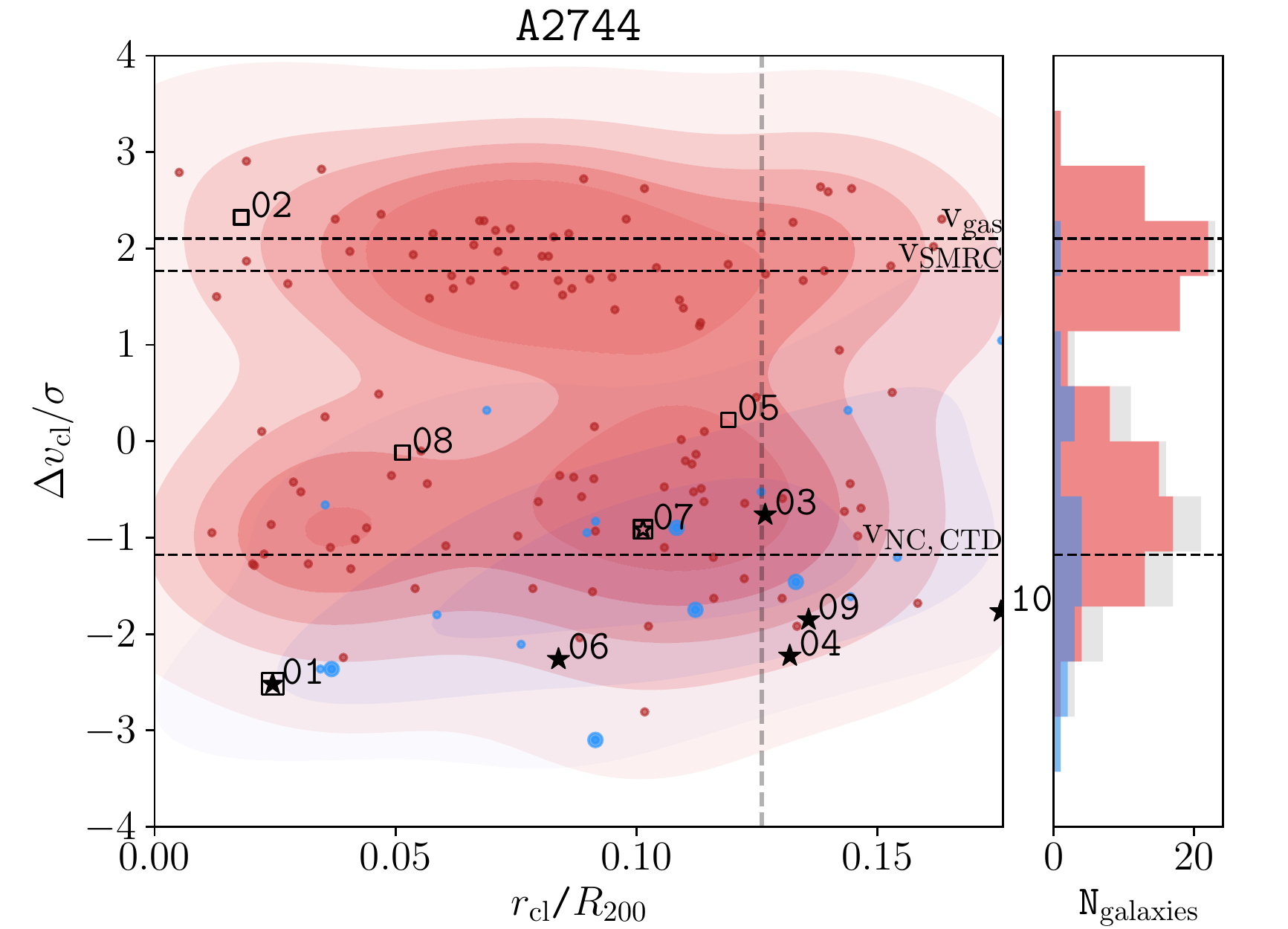}
    \includegraphics[width=0.49\textwidth]{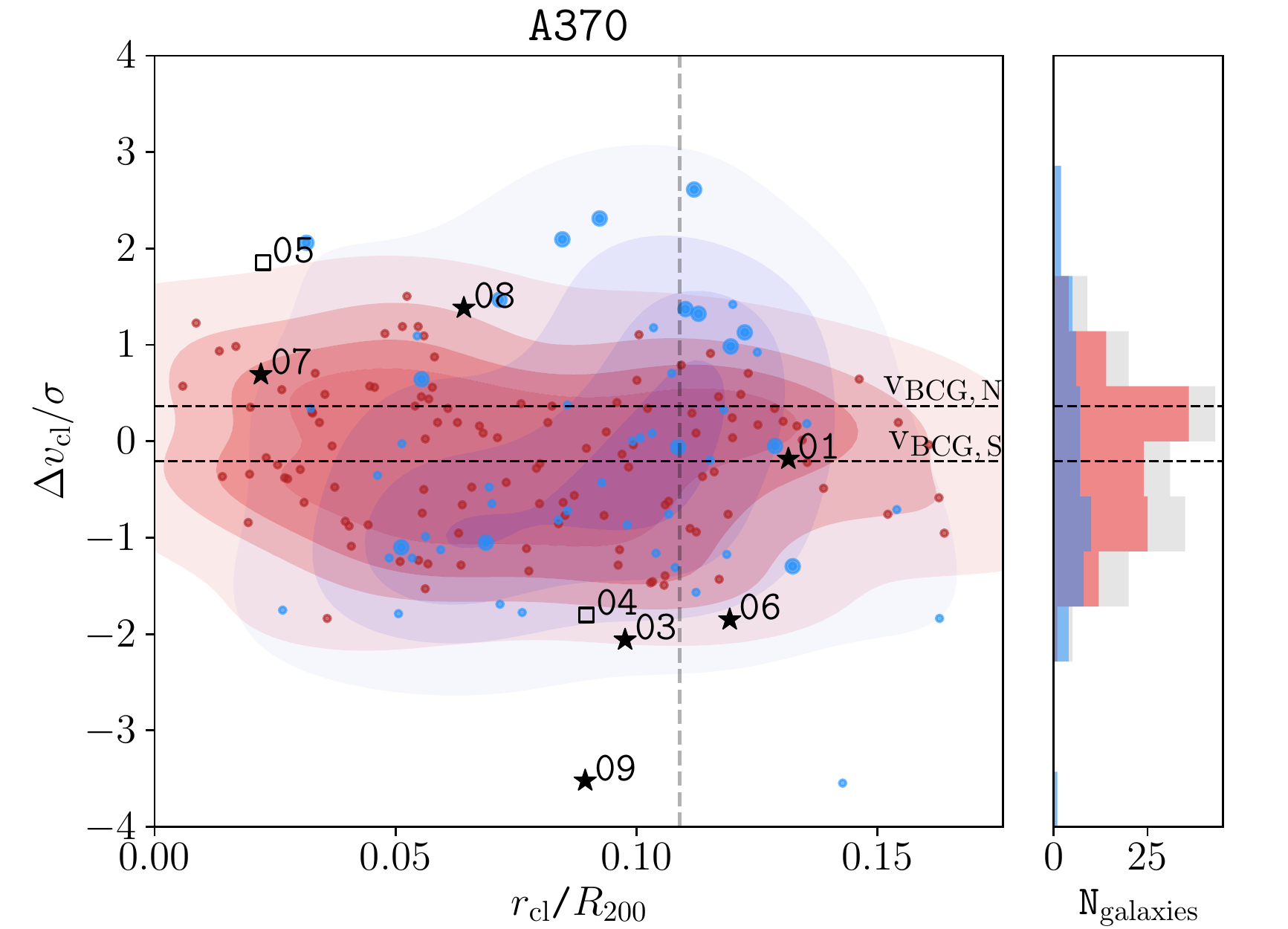}
    \caption{Phase-space diagram for the cluster member galaxies in the MUSE field of view of A2744 (\textit{left}) and A370 (\textit{right}). Galaxies within the color-magnitude red sequence are shown as red points. Galaxies below the red sequence are shown in blue, with galaxies bluer than 0.6 on the CM diagram highlighted as larger circles. The KDEs of the red and blue galaxy distributions are shown in respectively colored contours. Our sample of RPS galaxies is shown as black stars, whilst the PSB galaxies are shown as black squares. The vertical grey dashed line denotes the 100\% completeness radius described in the text. The vertical histograms show the velocity distributions of each cluster, with the same vertical axis as the phase space diagrams. The blue and red histograms show the blue and red populations accordingly, with the light gray histogram showing the combined total.
    \\\textit{A2744, Left:} Marked as dashed black lines are the reference velocities of the X-ray gas (v$_{\rm gas}$), the SMRC (v$_{\rm SMRC}$) and the velocity of both the NC and CTD (v$_{\rm NC,\, CTD}$), from \citet{Owers2011}. A2744\_01 and A2744\_07 are marked as both RPS and PSB, for the reasons described in section~\ref{sec:ident}. The cluster is clearly bimodal in velocity distribution from the histogram, and the majority of blue galaxies and RPS galaxies are located in the blueshifted component.
    \\\textit{A370, Right:} Black dashed lines indicate the reference velocities of the northern and southern BCGs, from \citep{Lagattuta2019}, as explained in the text. From the velocity histogram, the cluster appears to have a comparably singular velocity distribution in contrast to A2744 and the two merging components do not appear to be significantly separated. Furthermore, the blue galaxies are distributed across a wider range of velocities compared with the red galaxies. The RPS galaxies, PSB galaxies and the majority of the bluest galaxies with F606W-F814W$<0.6$ are generally situated at extremes in velocity with respect to the observed component of the cluster.
    }
    \label{fig:PS_both}
\end{figure*}

For the center of A2744, we use the coordinates of the BCG closest to the X-ray peak, as used in \citet{Owers2011}, and an R$_{200}$ radius of 2.00 Mpc measured by \citet{Boschin2006}. For each cluster we plot the projected radial distance from the center relative to R$_{200}$ and the line-of-sight velocity deviation from the cluster average, relative to the cluster global velocity dispersion, for all cluster galaxies. The resulting phase-space diagrams are shown in Figure~\ref{fig:PS_both}. The galaxies are divided into red and blue (and marked with accordingly colored points) based on the red sequence classification described in section \ref{sec:membership}. Red and blue filled contours show the kernel density estimates (hereafter KDE) of their respective galaxy samples, to better visualize their distribution within phase space. We mark the galaxies in our RPS sample with solid black stars, and the PSB galaxies with open black squares. In both figures, we denote a 100\% azimuthal completeness radius with a vertical dashed gray line. A circular aperture larger than this radius begins to extend beyond the boundaries of the square field-of-view of MUSE, therefore limiting the number count of visible galaxies. We also mark the average velocities of three important regions described in section \ref{sec:clusters}, measured by \citet{Owers2011}, which are the NC, the CTD and SMRC, as well as the X-ray velocity from the same paper, of the region therein described as MISC2.

The phase space diagram for A2744 shows the extremely disturbed, non-virialized nature of the cluster. Two distinct clumps are visible with a velocity separation of approximately $3\sigma$. The clump at $-1\sigma$ contains all galaxies in the RPS sample, as well as a much higher fraction of blue galaxies. 
The difference in the blue fraction between the two components could be the result of increased star formation \citep{Vulcani2018} in the CTD galaxies due to mild interaction with the ICM \citep{Stroe2017, Stroe2020}, or a higher quenched fraction in the SMRC as a result of its previous merging history.

In addition to this, some of our RPS galaxies are located within the MISC2 region identified in \citet{Owers2011}, corresponding to the X-ray surface brightness peak, shown in Figure~\ref{fig:A2744_footprint} in Appendix~\ref{sec:appendix}. In \citet{Owers2011}, a value of $0.3189^{+0.0092}_{-0.0110}$ was measured for the redshift of the X-ray emission, yielding a peculiar velocity of 2854.78$\mathrm{km\,s^{-1}}$ (marked v$_{\rm gas}$ in the left panel of Figure~\ref{fig:PS_both}). This velocity is very similar to the velocity of galaxies in the SMRC (marked v$_{\rm SMRC}$ in the same figure), and greatly offset from our sample of RPS and PSB galaxies.

The measured velocities of the RPS sample, as well as the velocities of the merging components of the cluster and the X-ray emitting gas, suggest that the RPS galaxies and most of the PSB galaxies are associated with the NC/CTD, but are passing through a region of gas which is moving with the SMRC. The extreme difference in velocity between the galaxies in this region and the cospatial gas is likely to be inducing the ram-pressure effect observed in our sample of RPS galaxies. This also corroborates the hypothesis put forward by \citet{Owers2011} that the CTD region consists of tidal debris removed from the main cluster (now the NC) during the core-passage phase of its merger with the SMRC.

In the case of A370, the phase space diagram shows a comparatively more relaxed distribution of galaxies, despite A370 also being a merging cluster.
The observed portion of the cluster has a skewed but single-peaked redshift distribution, as seen in the right-hand panel of Figure~\ref{fig:PS_both}. The velocities of the northern and southern BCGs are indicated as black dashed lines on the figure. Previous studies \citep{Lagattuta2019, Molnar2020} have noted that there are no prominent subgroups associated with the velocities of the BCGs. Instead, the velocities of the cluster galaxies appear to follow a more uniform Gaussian distribution.

The blue galaxies in A370 are generally located at higher clustercentric radii (median clustercentric distance: $0.25\, \mathrm{Mpc}$, $0.10\,\mathrm{R}_{200}$) compared with the red galaxies (median clustercentric distance: $0.19\,\mathrm{Mpc}$, $0.07\,\mathrm{R}_{200}$) and the velocity dispersion of the blue galaxies ($\sigma_{\rm blue}=2251\mathrm{km\,s^{-1}}$) is higher than that of the red galaxies  ($\sigma_{\rm red}=1320\mathrm{km\,s^{-1}}$). The velocity distribution of blue galaxies also appears to be slightly skewed towards negative values according to the histogram, although not as prominently as in A2744. 2 sample K-S tests reveal that at the 10\% significance level, the two populations follow the same velocity distribution ($p=0.14$) but distinct distributions in clustercentric radius ($p=0.05$).
The RPS and PSB galaxies in our sample are generally located at high positive and negative line-of-sight velocities and are not restricted to any particular region, but are distributed throughout the cluster. The high LOS velocities of the galaxies are conducive to ram-pressure due to the velocity difference between the galaxies and the ICM, whilst the scatter in locations suggests that infall is the root cause, in contrast to a large-scale movement or cluster interaction, which would affect galaxies in a specific region as we observe in A2744.

We therefore find two different stories regarding the cause of RPS between these two clusters. In the case of A2744, the RPS galaxies are likely to be experiencing an intense interaction with the ICM in the CTD region, directly resulting from the merging activity of the cluster. The galaxies, likely part of the CTD stripped from the NC, are colliding with a dense region of the ICM associated with the SMRC, which has a significantly different velocity. Similar scenarios of merger-induced RPS have previously been discussed in other clusters \citep{Ebeling2019, Stroe2020}. In contrast, within the observed region of A370, the RPS galaxies appear to be isolated infallers, experiencing ram-pressure as they accelerate into the cluster potential well. If, as discussed in \citet{Lagattuta2019}, the cluster has undergone an initial passage, it is possible that the disturbance of the ICM is enhancing the ram-pressure, but we do not observe a consistent, large-scale motion as in A2744.

\subsection{Cluster Substructure Analysis}

We carry out Dressler-Shectman (hereafter DS) \citep{Dressler1980,Knebe2000} tests to investigate where our galaxy samples are located within the context of the clusters' substructures. The DS test compares the velocity distribution of each galaxy and its 10 nearest neighbors to the velocity distribution of the cluster to identify regions of consistent velocity that are significantly offset from the cluster in general, indicative of a group or substructure.

For each galaxy and its 10 nearest neighbors, we measure the group standard deviation $\sigma_{\rm g}$ using the gapper method \citep{Beers1990} taking the differences between the sorted velocities of the galaxies and weighting by an approximately Gaussian envelope.

The deviation of each galaxy and its nearby companions is then defined as:

\begin{math}
\delta^{2} = \frac{11}{\sigma_{\rm cl}^{2}}\left[\left(\Bar{v}_{\rm g, pec}\right)^{2}+\left(\sigma_{\rm g} - \sigma_{\rm cl}\right)^{2}\right]
\end{math}

where $\sigma_{\rm cl}$ is the cluster standard deviation measured from the literature, $\sigma_{\rm g}$ is the gapper method standard deviation of the group and  $\Bar{v}_{\rm g,pec}$ is the group mean peculiar velocity, $\Bar{v}_{\rm g,pec} = \Bar{v}_{\rm g} - \Bar{v}_{\rm cl}$.

The value of $\delta$ gives a measure of the local deviation in velocity from the cluster as a whole. Groups of galaxies with similar velocities that are significantly different to the cluster average will have higher deviations, highlighting possible regions that kinematically stand out from the cluster as a whole.

In the case of A2744, \citet{Owers2011} consider the system to be a post-core merger, and due to the recent core passage, the main structures are still not fully disrupted and are clearly separated in velocity (as shown in the left panel of Figure~\ref{fig:PS_both}).
In this work, we focus on the smaller FoV of the MUSE observations, which exclude the NC but cover the SMRC and CTD structures (see Figure~\ref{fig:A2744_footprint} in Appendix~\ref{sec:appendix}). To better separate the SMRC and CTD structures, we perform the DS analysis considering the SMRC as the main structure, which naturally highlights the deviation of galaxies potentially belonging to the CTD (see left panel of Figure~\ref{fig:PS_both}). We note, additionally, that we also tried performing the DS test using the average velocity and velocity dispersion of all the cluster galaxies measured by \citet{Owers2011} and found signs of deviation in both groups, confirming this result (not shown).

\begin{figure*}
    \centering
    \includegraphics[width=0.49\textwidth]{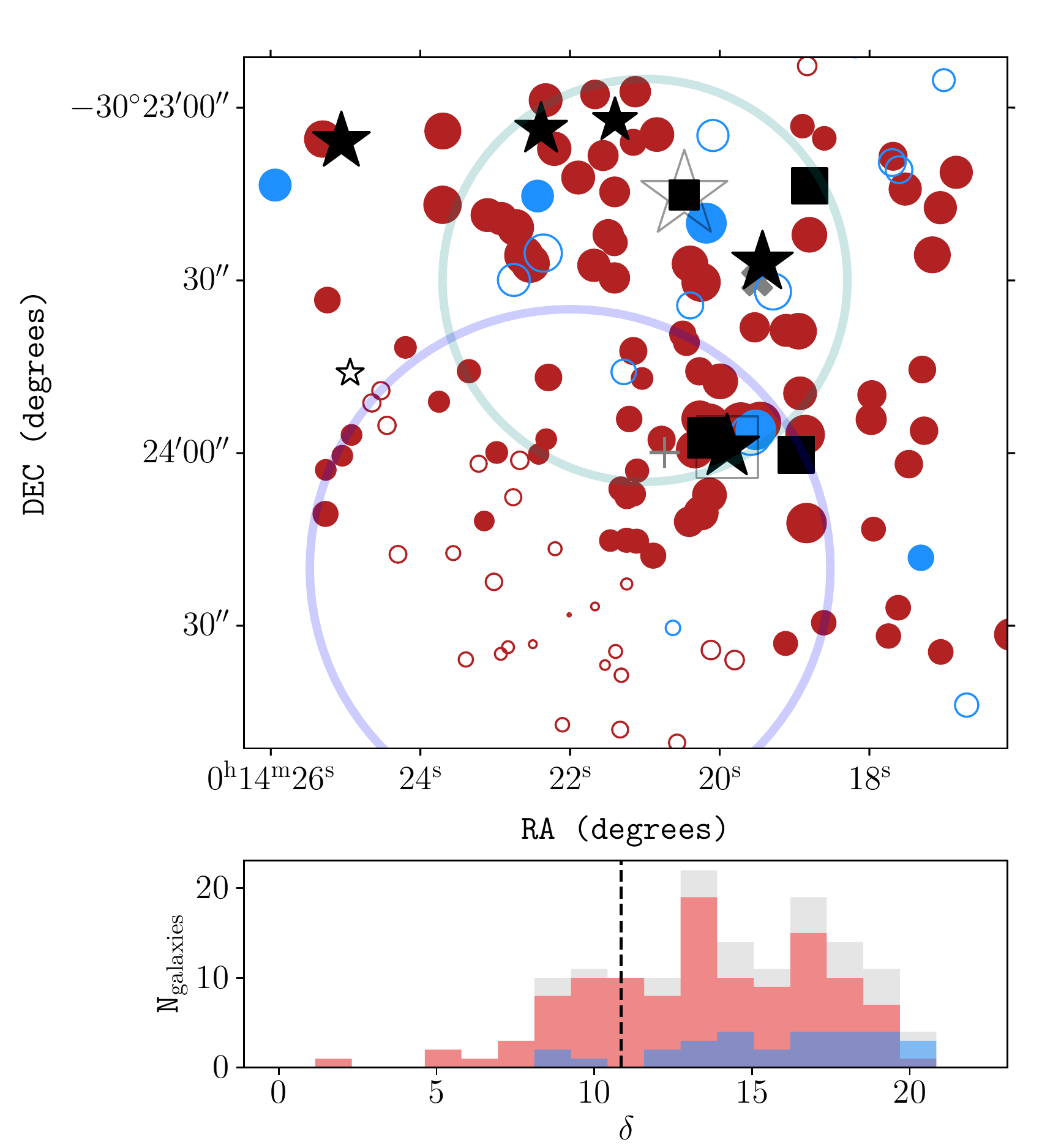}
    \includegraphics[width=0.49\textwidth]{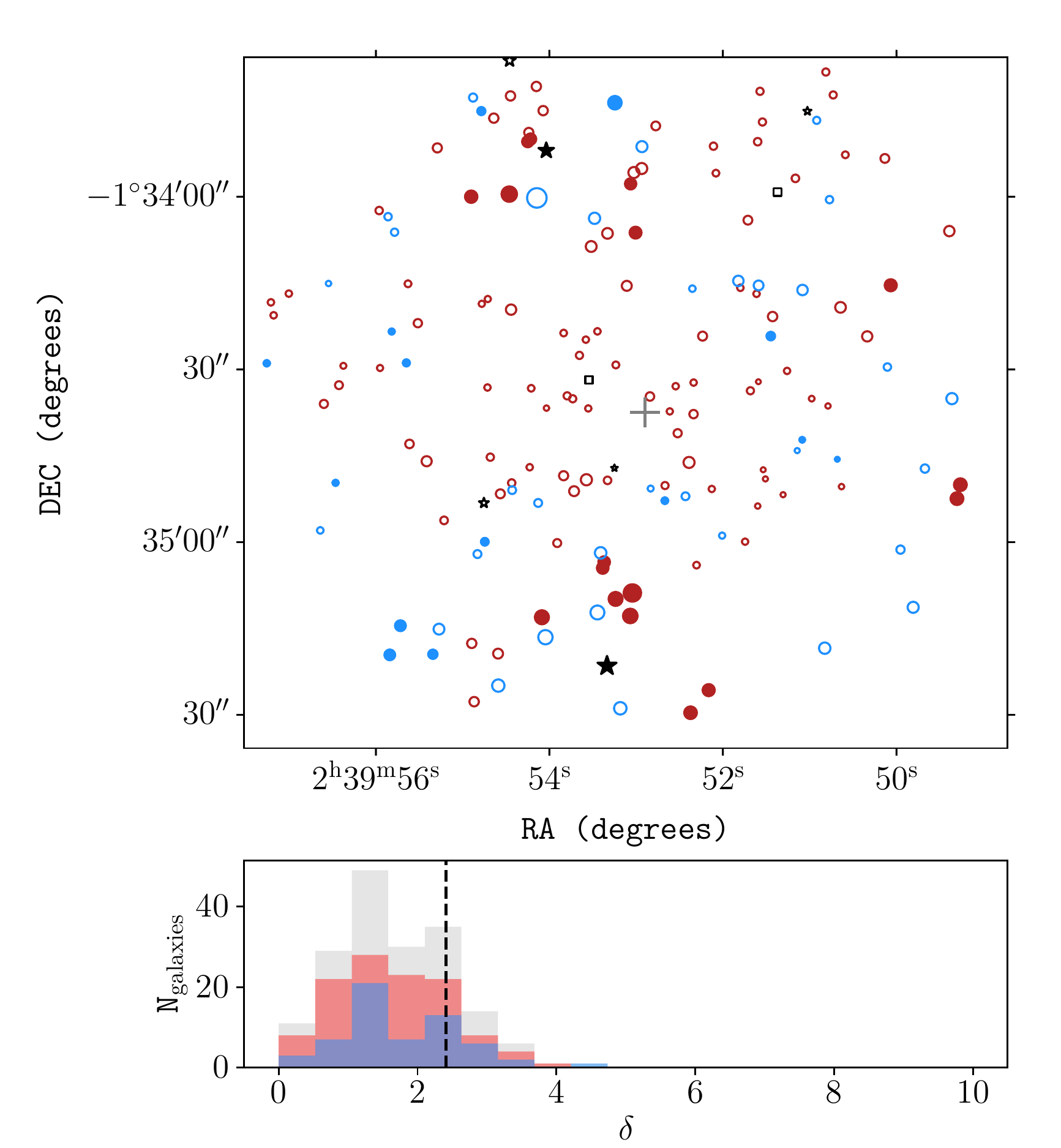}
    \caption{DS tests for A2744 (\textit{left}) and A370 (\textit{right}), showing bubble plot in upper panel and histogram of delta distribution in lower panel. Galaxies are shown as circles, highlighted red if they lie on the red sequence described in section \ref{sec:membership}, or blue otherwise. Black stars indicate the ram-pressure stripped sample and black squares indicate the PSB sample. A2744\_01 is highlighted as a black star in a light grey square, whilst A2744\_07 is highlighted as a black square in a light grey star, for the reasons explained in section \ref{sec:ident}. All points on the figure are filled if their velocity deviation surpasses the substructure threshold described in the text (black dashed line in lower panel), or unfilled otherwise. The cluster centers used in this analysis are marked with + symbols. The histograms below each plot show the distribution of delta values from the DS test for blue and red galaxies, colored accordingly, with the total combining both colors shown in gray. Black dashed vertical lines indicate the threshold for substructure as described in the text. For A2744, we mark on the bubble plot the regions of interest defined in \citep{Owers2011}: The light green circle marks the CTD, the blue circle marks the SMRC, and the gray X marks the location of the X-ray surface brightness peak.}
    \label{fig:DS_test_both}
    
\end{figure*}

The DS test results are shown in Figure~\ref{fig:DS_test_both} for both clusters. Galaxies in the clusters are shown as circles, with the size corresponding to their $\delta$ value. The size scale is enhanced $10\times$ in the case of A370 for clarity. Close groups of large circles are indicative of regions of substructure, where several galaxies are moving with a significantly deviated velocity from the cluster average. The limiting threshold on the delta value, indicating a significant deviation from the cluster velocity, is calculated to be $3\times\sigma_{\delta}$, where $\sigma_{\delta}$ is the standard deviation of all velocity deviation values in the given cluster \citep{Girardi1996, OlaveRojas2018}. Galaxies with a $\delta$ above this limit are shown with filled points, while those below the limit are shown unfilled. The RPS galaxies are marked with stars and the PSB galaxies are marked squares. A2744\_01 is marked as both RPS and PSB for the reasons outlined in section \ref{sec:ident}.

The DS test for A2744 indicates that many galaxies are moving with velocities which are significantly deviated from the SMRC. This is expected, considering the cluster is undergoing a significant merging event, and the majority of galaxies in the field of view are associated with different merging components. Our DS test for this region of the cluster concurs with the results of \citet{Owers2011}, the most significant substructure in the field of view is the CTD, visible across the north west half of the figure as a grouping of large circles. All of the ram-pressure stripped galaxies and PSB galaxies in our sample are parts of some velocity substructure, which is also expected if the merging of the different components is driving the onset of stripping.

Combining these results with the phase space analysis, we find that several pieces of evidence consolidate the hypothesis that these galaxies are experiencing a magnified interaction with the ICM resulting from the merger event:

\begin{enumerate}
    \item The presence of many RPS and PSB galaxies in the velocity component at $-1\sigma$, along with the difference in the blue galaxy fraction compared with the component at $2\sigma$.
    \item The location of these galaxies within or close to the region postulated by \citet{Owers2011} as the CTD (shown by the grey circle in Figure~\ref{fig:DS_test_both} and the green circle in Figure~\ref{fig:A2744_footprint} in Appendix~\ref{sec:appendix}), resulting from the major merger of the clusters.
    \item The velocity of these galaxies being similar to the average velocity of galaxies in the CTD and the NC (marked v$_{\rm NC}$ in the left hand panel of Figure~\ref{fig:PS_both}, measured by \citet{Owers2011}.
\end{enumerate}

The DS test for A370 highlights a more relaxed distribution of galaxies within the observed region of the cluster in comparison to A2744. The majority of galaxies within the MUSE field of view are not indicated to reside within any separate substructures, and only a few significant substructures are detected, to the north east, south west, and to a lesser extent, the west of the figure. All but two of the RPS galaxies and all of the PSB galaxies are not located within any of the detected substructures, which may suggest that the majority of our sample entered the cluster as isolated galaxies.

\section{Comparison with X-ray and Gravitational Lensing analysis}\label{sec:xray+lensing}

\subsection{X-ray and mass surface density maps}

\begin{figure*}
    \centering
    \includegraphics[width=\textwidth]{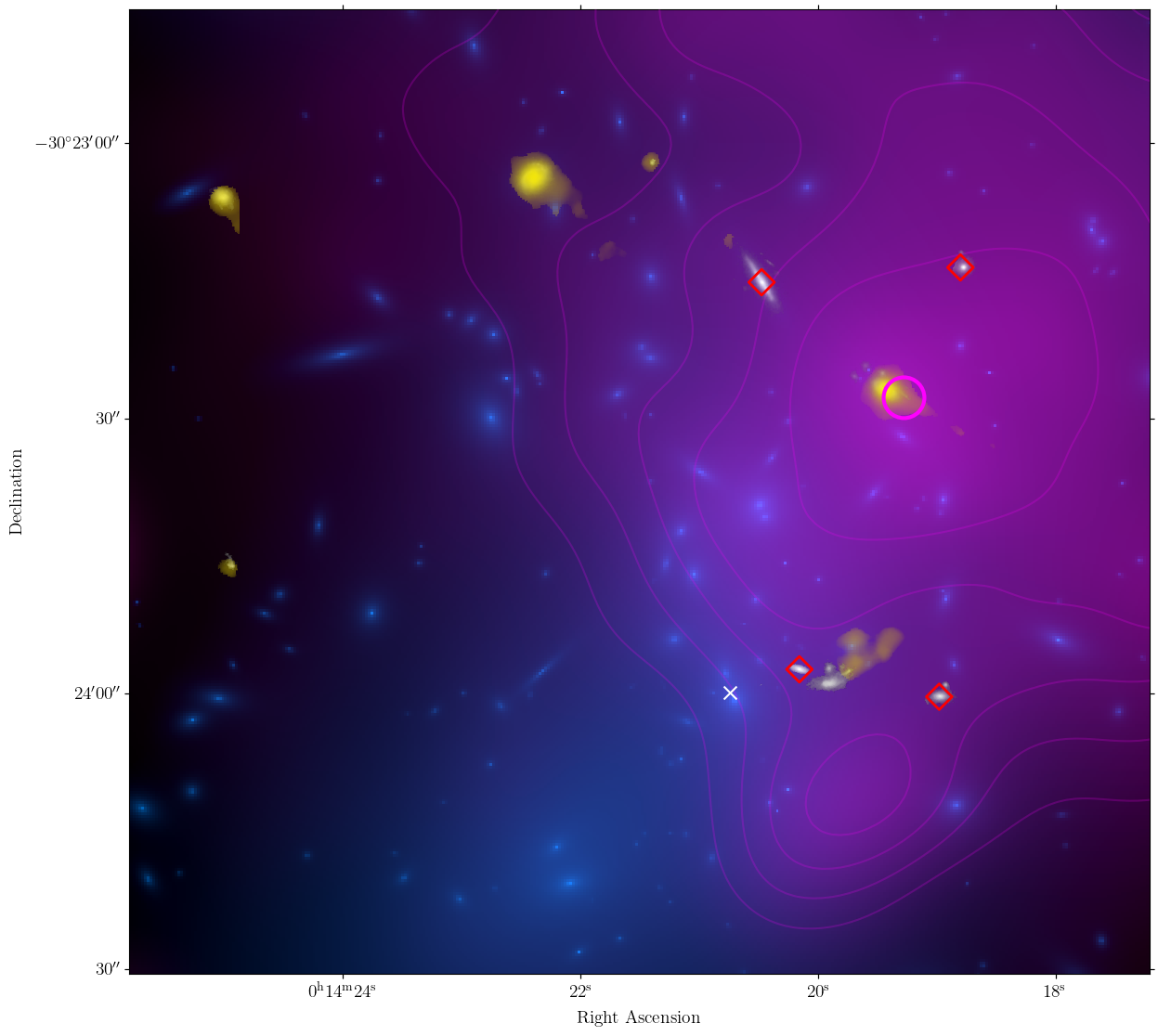}
    \caption{X-Ray and Lensing maps of A2744 with F606W + H$\alpha$ maps of RPS galaxies and PSBs overlaid. The magenta image and contours show the smoothed X-ray map from Chandra. The CATS v4 lensing mass map is shown in cyan. F606W images of RPS and PSB galaxies are shown in white, with MUSE H$\alpha$ maps shown in yellow. PSB galaxies are marked with red diamonds. The cluster center used in the analysis is marked with a white X. The peak in the X-ray emission is marked with a magenta circle.}
    \label{fig:A2744_overview}
\end{figure*}

\begin{figure*}
    \centering
    \includegraphics[width=\textwidth]{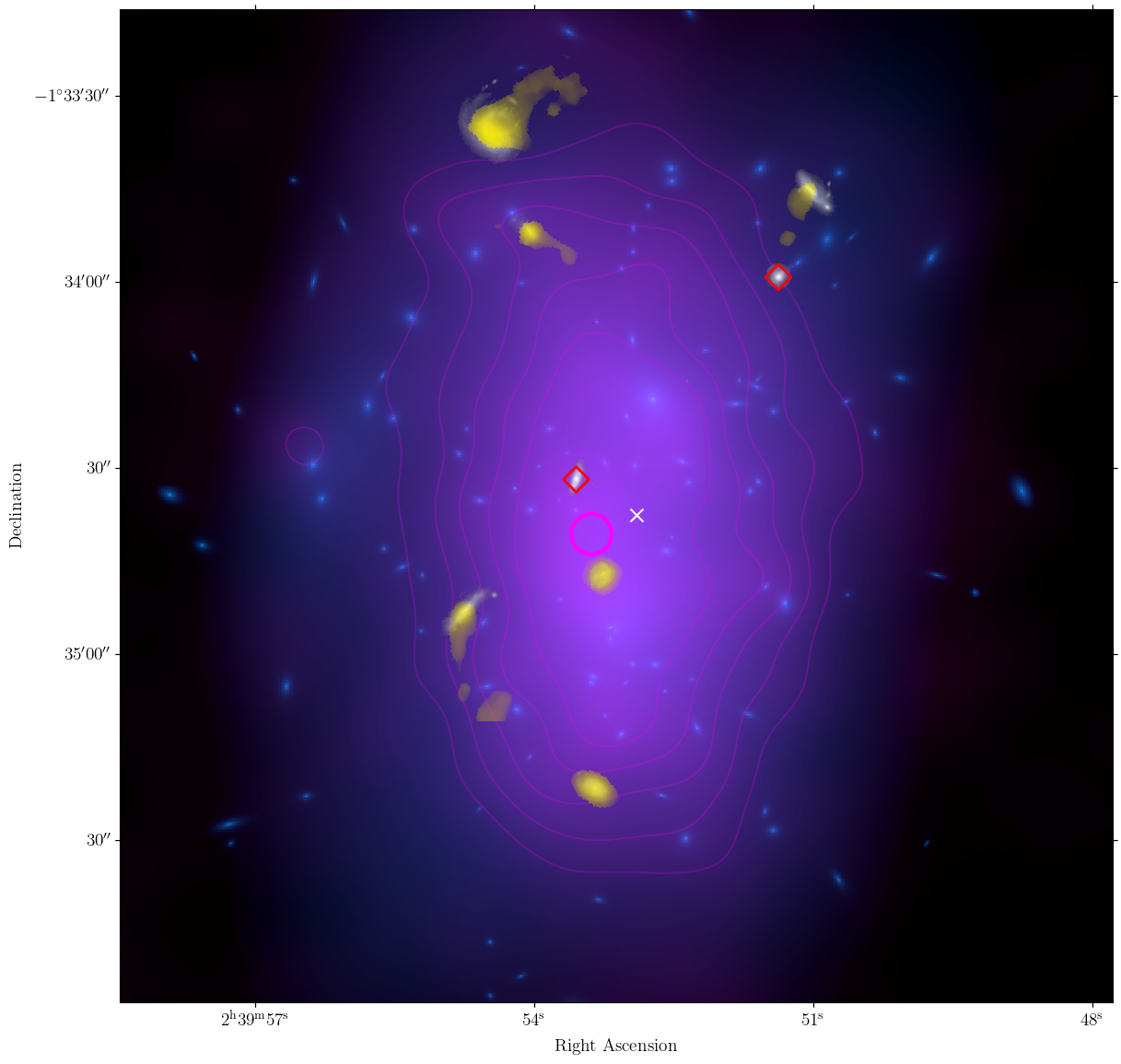}
    \caption{X-Ray and Lensing maps of A370 with F606W + H$\alpha$ maps of RPS galaxies and PSBs overlaid. The magenta image and contours show the smoothed X-ray map from Chandra. The CATS v4 lensing mass map is shown in cyan. Whitelight images of RPS and PSB galaxies are shown in white, with MUSE H$\alpha$ maps shown in yellow. PSB galaxies are marked with red diamonds. The cluster center used in the analysis is marked with a white X. The peak in the X-ray emission is marked with a magenta circle.}
    \label{fig:A370_overview}
\end{figure*}

In order to investigate the distribution of RPS and PSB galaxies within the environment of the cluster, we plotted the locations of the galaxies overlaid on the X-ray and the lensing modelled mass surface density maps, shown in Figures \ref{fig:A2744_overview} and \ref{fig:A370_overview} for A2744 and A370 respectively. For both clusters, we show the CATS version 4 \citep{Lagattuta2017, Lagattuta2019, Mahler2018} mass surface density map in cyan and the Chandra X-ray image  \citep{Mantz2010,vonderLinden2014,Vulcani2017} in magenta. For display purposes, the X-ray images were smoothed with a $3\times3$ median filter and convolved with a $5\times5$ Gaussian kernel.
For each of the RPS and PSB galaxies, we plot the cleaned HST F606W map in white, as well as the MUSE H$\alpha$ map shown in yellow. The H$\alpha$ map was produced using our custom-made emission-line fitting software \textsc{highelf} (Radovich et al. \textit{in preparation}) which is based on the \href{https://lmfit-py.readthedocs.io/}{LMFIT python library (https://lmfit-py.readthedocs.io/)} and fits a user-defined set of emission lines using one or two Gaussian components (see \citealt{Moretti2022}, Section~3 therein for full details of the measurement). The white x marks in each figure show the centers of the clusters used in this study.

The cluster maps highlight the differences between the observed regions of the clusters. A2744 has a prominently disturbed gas component, shown by the X-ray map, with the majority of the X-ray emitting gas located to the upper right of the observed region. The mass component, shown by the lensing modelled mass surface density map, is distinct from the X-ray component and the majority of the mass is located to the lower left of the X-ray emission. This large difference between the galaxies and gas is attributed in \citet{Owers2011} to the collision between the major components of the cluster which has decoupled the gas component from the collisionless galaxies and dark matter components.

In comparison, the X-ray and mass surface density maps for A370 are coincident, with the peak of the X-ray emission lying between the two peaks of the mass map (see also \citealt{Lagattuta2017}, Figure~10).

In the case of A2744, as discussed in section \ref{sec:phasespace}, the majority of the PSB and RPS galaxies in our sample are located within the X-ray region, to the upper and/or right of the majority of the mass distribution. This X-ray emitting gas has a high velocity offset from the RPS and PSB galaxies. Since this velocity offset is the result of the merger between this subcluster and the NC, the merging activity appears to be driving the RPS interactions, rather than infall alone.
In particular, the RPS is being enhanced by the collision between the galaxies associated with one merging component, and the gas associated with another merging component. One galaxy of interest in this cluster is A2744\_01, which was classified both as a RPS and PSB galaxy, due to its long tail of stripped material but lack of star formation or nebular emission lines. The direction of the galaxy's tail suggests that it has recently passed through the region characterized by very strong ICM X-ray emission, indicating that the galaxy may have just exited a phase of very strong RPS.

For A370, the RPS and PSB galaxies are more uniformly distributed around the cluster, and not constrained to a particular region. Together with the more uniform distribution of the X-ray component and its alignment with the mass surface density distribution, this suggests that the galaxies are undergoing stripping due to infall rather than collision with a merging component's ICM as in A2744.




\subsection{Comparing ICM X-ray emission}

\begin{figure}
    \centering
    \includegraphics[width=0.5\textwidth]{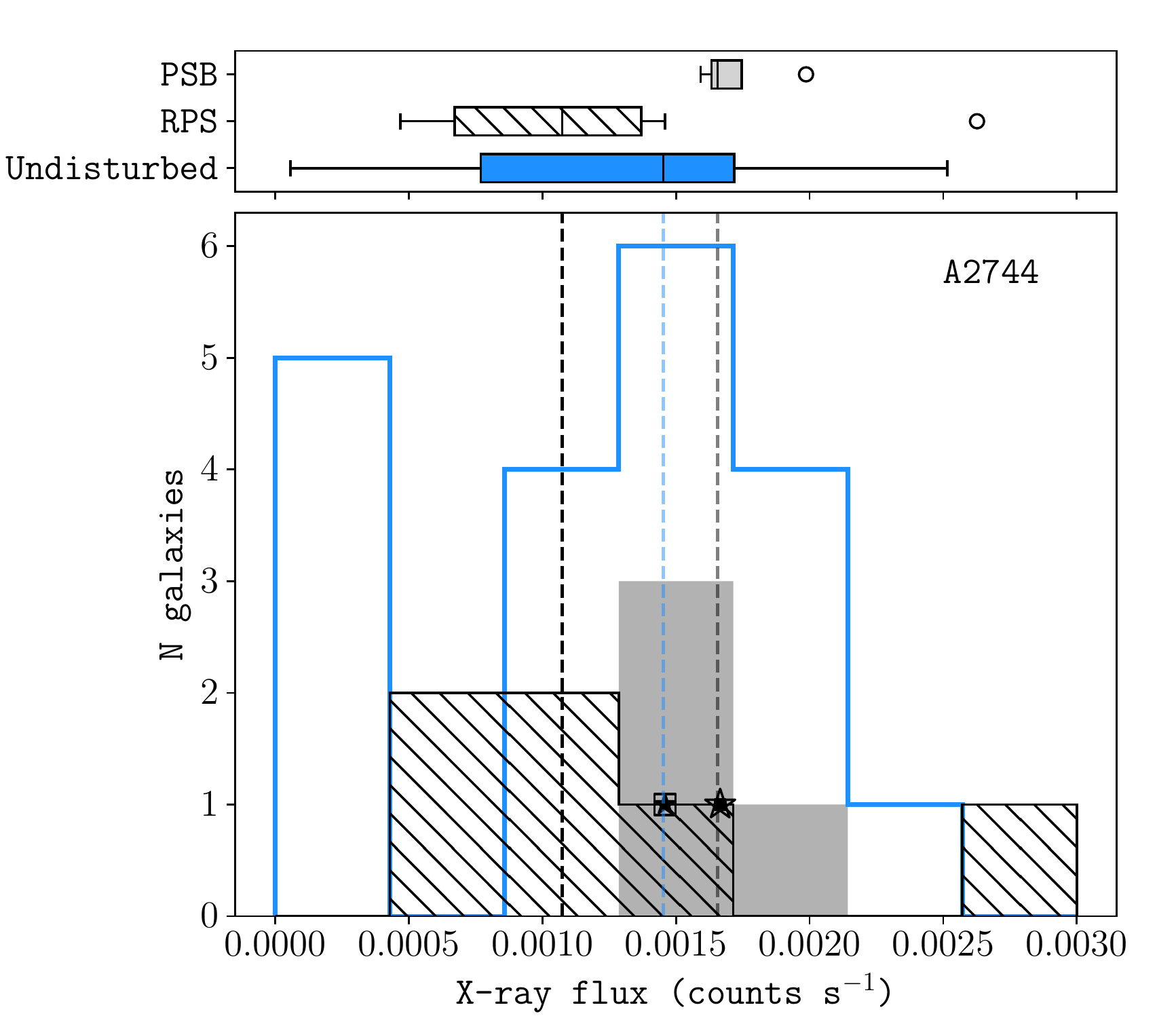}
    \includegraphics[width=0.5\textwidth]{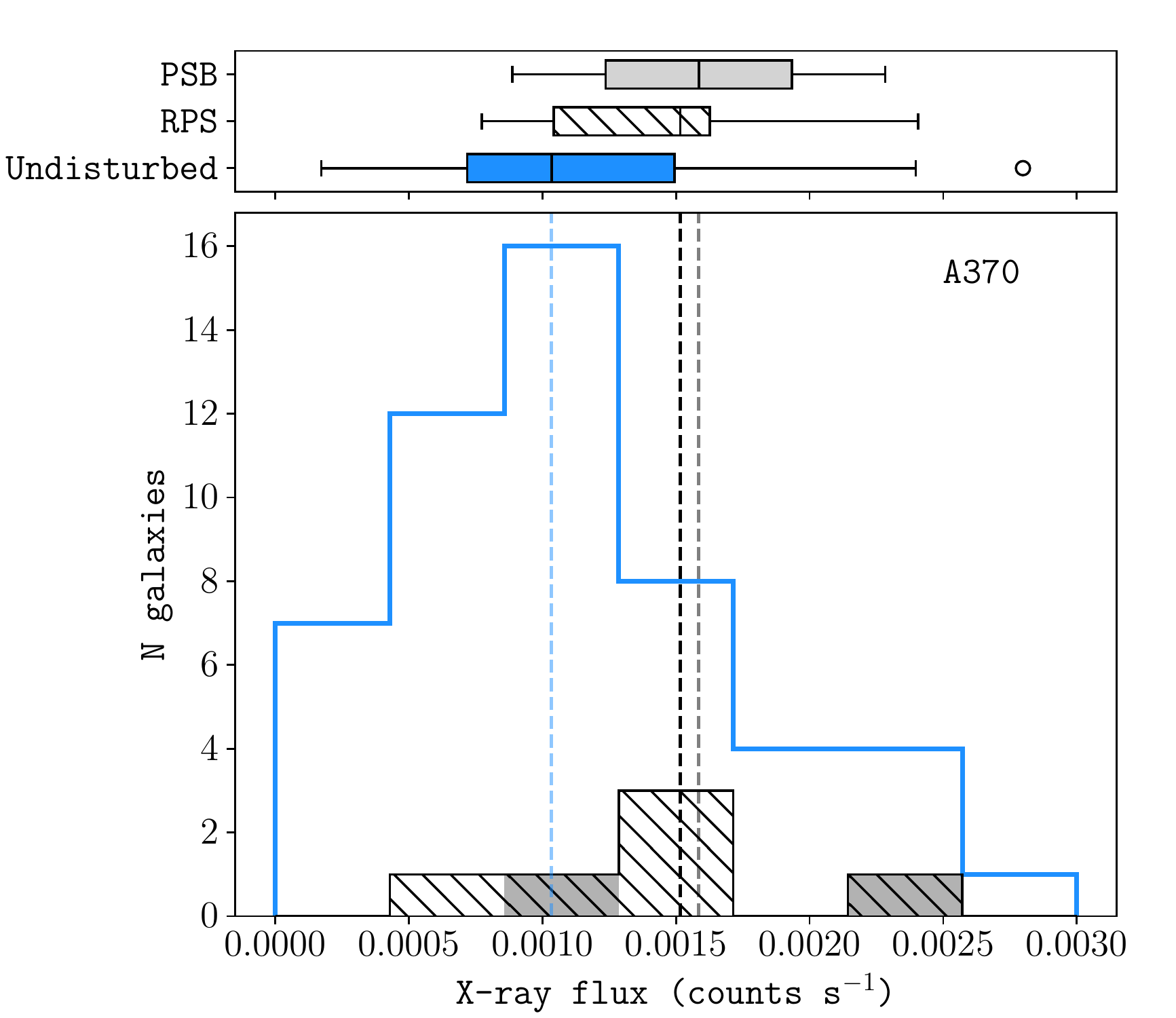}
    \caption{Histogram of coincident ICM X-ray fluxes at the locations of all PSB galaxies (grey shaded) all RPS galaxies (black hashed) and all remaining blue cluster galaxies (blue) for A2744 (top) and A370 (bottom). The median values for the ICM X-ray flux for each sample are plotted as dashed vertical lines with the same colours. The location of A2744\_01 is highlighted with a black star in a hollow box, whilst the location of A2744\_07 is highlighted as a black box in a hollow star, for the reasons explained in section \ref{sec:ident}.}
    \label{fig:xray_hist}
\end{figure}

\begin{figure}
    \centering
    \includegraphics[width=0.5\textwidth]{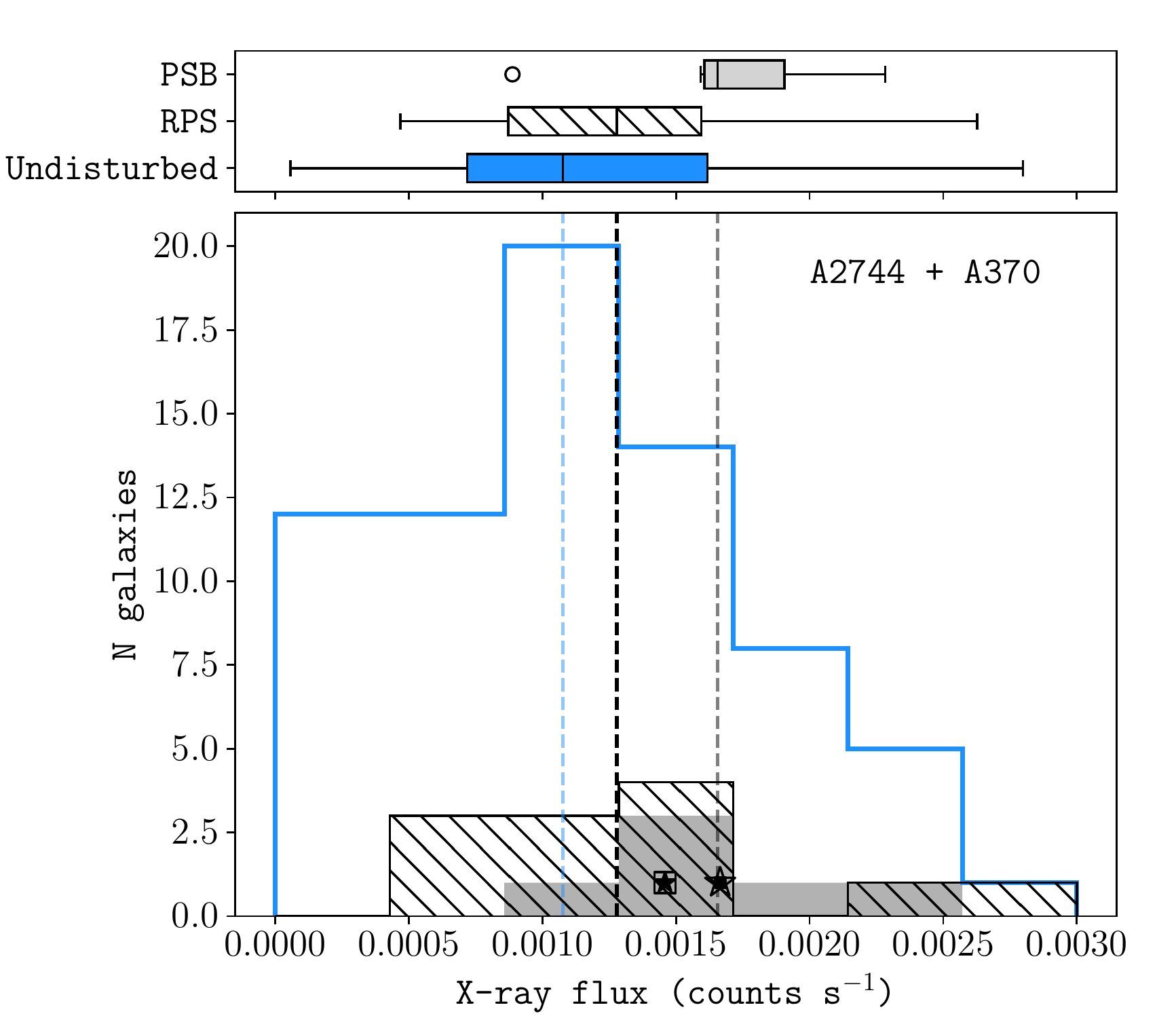}
    \caption{Histogram of coincident ICM X-ray fluxes at the locations of all PSB galaxies (grey shaded) all RPS galaxies (black hashed) and all remaining blue cluster galaxies (blue) for A2744 and A370 combined. The lines and symbols are as described in Figure~\ref{fig:xray_hist}.}
    \label{fig:xray_hist_combined}
\end{figure}

Several works have explored the correlation between stripping efficiency and the presence of X-ray gradients and shocks from the ICM \citep{Owers2012,Vijayaraghavan2013}. \citet{Vulcani2017} observed that in unrelaxed clusters, H$\alpha$ emitter properties exhibit slight trends with the local ICM X-ray emission, suggesting that some form of interaction with these features may be responsible for the stripping of gas and/or triggering of star formation. In addition, \citet{Stroe2020} find strong evidence for triggering of SF by shocks produced by merging activity in the post-core passage merging cluster CIZA J2242.8+5301, nicknamed the Sausage cluster. The alignment of disturbed features in the Sausage cluster galaxies with the merger axis of the cluster strongly indicates that the galaxies have been disturbed by interactions with the travelling shock fronts in the ICM.
In order to investigate whether we see a correlation between stripping activity and ICM X-ray emission, we investigated the coincident X-ray flux of the ICM at the locations of the galaxies in our sample with the cluster members. To do this we calculated the median X-ray flux within an annulus, avoiding any X-ray emission from the galaxies themselves, between 15 and 30 kpc from the location of each galaxy. The X-ray fluxes are shown in Figure~\ref{fig:xray_hist} for A2744 (\textit{top}) and A370 (\textit{bottom}) with the RPS galaxies marked in black and white hatching and PSB galaxies marked in grey. The general population of blue cluster galaxies is shown in blue for comparison. We tentatively observe in both clusters that PSB galaxies are generally found within regions of higher X-ray flux emission from the ICM, in comparison to blue cluster galaxies. In the case of A370, the population of RPS galaxies are also found in regions of higher ICM X-ray emission compared with blue cluster galaxies, whilst in A2744 the distribution of ICM X-ray emission is comparable to, or lower than that of the blue cluster galaxies.
The differences between the samples are small, however. 2-sample K-S tests revealed no distinction at the 5\% significance level between the distributions of the RPS and PSB galaxies compared with the blue cluster galaxies in both A2744 ($p_{\rm RPS}=0.58$, $p_{\rm PSB}=0.27$) and A370 ($p_{\rm RPS}=0.28$, $p_{\rm PSB}=0.71$) individually. On the other hand, when the two clusters are combined as shown in Figure~\ref{fig:xray_hist_combined}, the distribution of PSBs becomes distinct from the blue cluster galaxies ($p_{\rm PSB}=0.04$).

In general, A2744 appears to be too disturbed to draw a reliable conclusion on its own. Compared with A370, the distribution of blue galaxies in A2744 appears to be pushed toward higher X-ray fluxes, which may be due to the dearth of blue galaxies in the SMRC (which corresponds a region of generally lower X-ray emission) resulting from its past minor merger.

Whilst we emphasise the caveat that the 3D distribution of the disturbed cluster ICM and the 3D locations of the galaxies are not known, our result for the PSB galaxies may be consistent with the progression of galaxies through the RPS stage as they pass through denser regions of the ICM into the PSB phase. Galaxies encountering interactions with denser regions of the ICM are expected to be subjected to more intense RPS, up to the point that the gas becomes fully stripped and the galaxy becomes a PSB galaxy. In this case, galaxies located in denser regions of the ICM may have already been stripped to the point of becoming PSB.

Our findings for the PSB galaxies may also be consistent with \citet{Vulcani2017}, who measured the offset between the peaks of the H$\alpha$ and F475W emission projected along the cluster radial direction, for cluster galaxies in the GLASS survey, and found that a correlation emerged with the ICM X-ray emission for galaxies in unrelaxed clusters.

\section{Morphology analysis}\label{sec:morphology}

We utilize several quantitative measures of the morphologies to understand whether our RPS and PSB galaxies occupy a specific region of morphology space. The selection of RPS and PSB galaxies, based the presence of H$\alpha$ tails measured from the MUSE data,
allows us to better explore their distribution in terms of visual morphology parameters, since the selection is not strictly biased towards galaxies with notable visual disturbances. Many galaxies in our sample appear fairly undisturbed in broad-band imaging, but have clear tails in the MUSE data. This allows us to test the sensitivity of these parameters in order to determine whether subtle cases of RPS can still be differentiated from the general cluster population using this analysis.

We also investigate whether this analysis is improved by the inclusion of multiple broad band filters, to incorporate color measurements, or if one broad-band filter is sufficient. This will determine the minimum requirements for potential sample selections based solely on these techniques.

For each galaxy, we make a cutout from the F606W HST image and use the python package \texttt{statmorph} \citep{Rodriguez-Gomez2019} to extract different morphological parameters. We ran the \texttt{statmorph} function \texttt{source\_morphology} using the segmentation maps produced by the GTO pipeline, with a gain of 2.5 and a point spread function (PSF) generated using \texttt{sextractor}. Weight maps and masks were not used in this case. The cutouts, segmentation maps and resulting morphology parameters are shown for a few examples in Figure~\ref{fig:segmentation} in Appendix~\ref{sec:appendix}.

We focus on the morphological quantities \textit{concentration} and \textit{asymmetry} \citep{conselice2003a, Conselice2014}, as well as \textit{gini} \citep{Glasser1962, Abraham2003,Lotz2004} and \textit{M$_{20}$} \citep{Lotz2004}. We note here brief summaries of the morphological parameters used and refer the reader to \citet{Roberts2020} for an effective summary and the original papers cited here for the full details.

The \textit{concentration} measure \citep{conselice2003a, Conselice2014} is derived from the ratio of the radii that contain 80 and 20 percent of the total luminosity of a galaxy, giving an indication of the steepness of the light profile of the source.

The \textit{asymmetry} \citep{conselice2003a, Conselice2014} parameter is defined as the difference between the flux map of a galaxy and its $180^\circ$ rotated counterpart. This parameter is particularly suitable for detecting the asymmetric offsets and disturbances expected during RPS interactions. We note also the \textit{shape asymmetry} \citep{Pawlik2016} parameter, utilized in \citet{Roberts2020}, which uses the binary detection map instead of the total flux map. This parameter is more sensitive to low surface-brightness features, making it ideal for detecting the disturbances visible in RPS, however the observations we use in this study have particularly crowded fields, impacting the shape of the detection maps. We found that the standard asymmetry, whilst less sensitive to low surface-brightness features, was a more robust measure in a crowded environment.

The \textit{gini} parameter \citep{Glasser1962} is traditionally used within the context of economics and parametrises the distribution of wealth across a population. The parameter has also been utilized in astronomy to quantify the distribution of flux in an image, with lower values indicating a more homogeneous distribution and higher values describing a more concentrated source of flux.

Finally, the \textit{M$_{20}$} statistic is the ratio of the second order moment of the brightest $20\%$ of pixels in a galaxy's image and the second order moment of the entire image. This parameter is sensitive to bright features offset from the galaxy center, which makes it suitable for detecting large disturbances in a galaxy's morphology.

\begin{figure*}
    \centering
    \includegraphics[width=\textwidth]{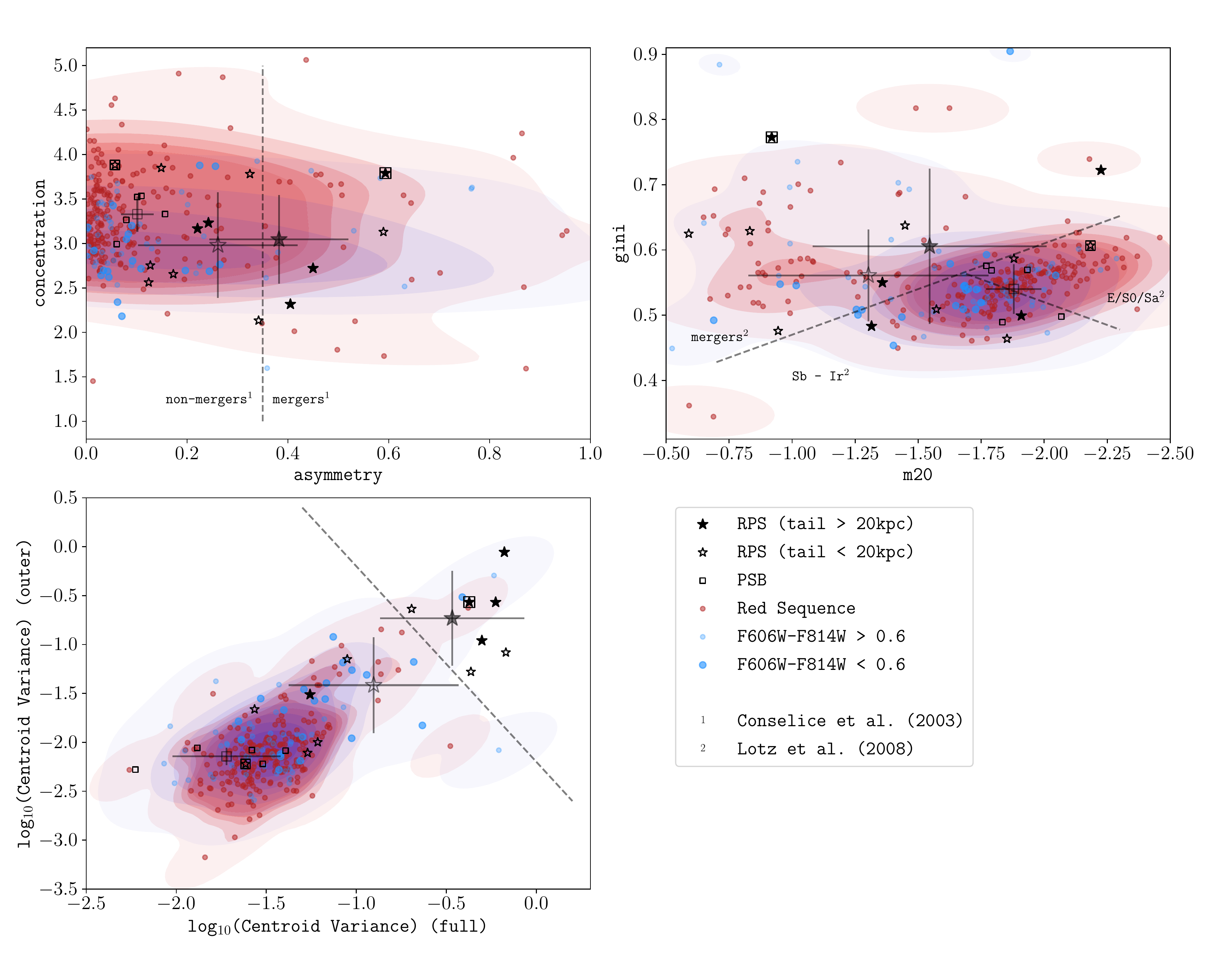}
    \caption{Morphology parameters for galaxies within A2744 and A370 combined. Galaxies are coloured accordingly as described in the colour magnitude diagrams, Figure~\ref{fig:CM_both}. RPS galaxies are marked with solid black stars if they have H$\alpha$ tails longer than 20kpc according to \citet{Moretti2022} and open black stars otherwise. PSB galaxies are marked with open black squares. A2744\_01 (black star in a hollow square) and A2744\_07 (hollow star in a hollow square) are marked thus as both RPS and PSB as explained in section \ref{sec:ident}.\\ \textit{Upper Left:} Plot of the \textit{concentration} and \textit{asymmetry} parameters, derived as described in the text. The median and the standard deviation of \textit{concentration} and \textit{asymmetry} within each category are shown as error bars marked with the relevant colour or symbol.\\
    \textit{Upper Right:} Plot of the \textit{gini} and \textit{M20} parameters, derived as described in the text. The median and the standard deviation of \textit{gini} and \textit{M20} within each category are shown as error bars marked with the relevant color or symbol.\\
    \textit{Lower Left:} Plot of the outer and full centroid variance parameters, derived as described in the text. The median and the standard deviation of the centroid variance within each category are shown as error bars marked with the relevant color or symbol.}
    \label{fig:morphology}
\end{figure*}

We combine the parameters \textit{concentration} and \textit{asymmetry} in the top-left panel of Figure~\ref{fig:morphology} for both A2744 and A370 together. In the figure, we mark red galaxies, classified as described in section~\ref{sec:membership} as red points, galaxies with F606W-F814W$<0.6$ as blue points, and intermediate colour galaxies as light blue points. The non-red cluster population is divided into two groups in this way to better highlight the locations of the bluest galaxies, in order to understand whether they have discernibly distinct morphologies according to this analysis. RPS galaxies are shown as solid black stars and PSB galaxies as open black squares. We also separate our RPS galaxy sample based on the lengths of the H$\alpha$ tails measured by \cite{Moretti2022}. Galaxies with tails longer than 20kpc are marked with solid black stars, and galaxies with tails shorter than this are marked with open black stars.

For the populations of red and blue galaxies, we show the KDE as correspondingly colored contours. For the PSB galaxies as well as the long and short tailed RPS galaxies, we plot the median and standard deviation as errorbars, marked with the relevant icon as appropriate. This is done in order to highlight the general distributions of each population in parameter space. We include on the figure the line of A=0.35 from \citet{conselice2003b}, which was given as a threshold to select merging galaxies. Whilst we do not classify merging galaxies in this study, it is useful to understand how our sample would be interpreted using this morphological criterion. We note that this threshold is subject to variation, as the measured asymmetry can vary with image resolution and depth \citep{Lotz2004,Sazonova2021,Thorp2021} and the \texttt{statmorph} asymmetries are lower than those used in \citet{conselice2003b} \citep{Rodriguez-Gomez2019}, but include the threshold for reference as an indicator of ``significant" asymmetry.

The parameters \textit{gini} and \textit{M$_{20}$} were combined in the top right panel of Figure~\ref{fig:morphology} for both A2744 and A370, with each sample marked and coloured as described above. We include in the figure the lines used in \citet{Lotz2008} to separate out different types of galaxies in \textit{gini}, \textit{M$_{20}$} space:

\begin{align*}
\mathrm{Mergers{:}}&~\mathrm{G} > -0.14 \mathrm{M}_{20} + 0.33;\\
\mathrm{E/S0/Sa{:}}&~\mathrm{G} \leq -0.14 \mathrm{M}_{20} + 0.33 ~\mathrm{and}~ \mathrm{G}>0.14\mathrm{M}_{20} + 0.80;\\
\mathrm{Sb - Ir{:}}&~\mathrm{G} \leq -0.14 \mathrm{M}_{20} + 0.33 ~\mathrm{and}~ \mathrm{G}\leq0.14\mathrm{M}_{20} + 0.80;
\end{align*}

As with the \textit{asymmetry} threshold, whilst we do not actively classify these types of galaxies in our sample, we include these lines to compare our sample with the classifications given by these criteria.


We find that the RPS galaxies are spread across a wide range of \textit{concentrations}, similarly to the rest of the cluster galaxies. The bluest galaxies, with F606W-F814W$<0.6$, appear to be less concentrated on average compared with the red galaxies. The \textit{asymmetry} provides a more prominent separation between our disturbed sample and the cluster galaxies, with the PSB galaxies located slightly higher than the average, and the RPS galaxies exhibiting much higher values compared with both the PSB galaxies and the rest of the cluster galaxies. The galaxies with long H$\alpha$ tails in the MUSE data have, on average, higher asymmetries in the broad-band data compared with those with shorter H$\alpha$ tails, suggesting that the visual disturbance in broad-band imaging is correlated with the underlying ionized gas disturbance. Comparing the sample with the $\mathrm{A}>0.35$ line, we find that the majority of our long-tailed RPS galaxies would be classified as mergers by this criterion, whilst most of the short-tailed RPS galaxies and PSB galaxies lie below the threshold. This suggests that whilst the $\mathrm{A}>0.35$ criterion is useful for selecting disturbed morphologies, it cannot be solely relied upon to distinguish the underlying cause of the disturbance without making considerations about the environment of the galaxies.

The \textit{gini} and \textit{M$_{20}$} parameters on their own do not strictly separate the RPS or PSB galaxies from the rest of the cluster sample, but when combined, we see that the bulk of the red cluster galaxies are located in a small region in the middle right of the figure,
whilst our sample of PSB galaxies are generally located around the outskirts of this region and the RPS galaxies occupy a much wider range of values, generally being found away from this concentrated region of cluster galaxies. The blue galaxies are fairly scattered in this figure, however there is a high concentration of galaxies with F606W-F814W$<0.6$ in the Sb-Ir region. The galaxies with short and long H$\alpha$ tails are not particularly differentiated by their \textit{gini} and \textit{M$_{20}$} values. We find that the mean values of the long and short tailed RPS galaxies lie within the merger region given by the \citet{Lotz2008} criteria. As with the asymmetry criterion, this implies that such criteria cannot distinguish between merging and RPS galaxies, and that samples of merging galaxies selected using these criteria may also include galaxies disturbed by ram-pressure.

In both the \textit{concentration}, \textit{asymmetry} and the \textit{gini}, \textit{M$_{20}$} figures, the galaxies A2744\_01 and A2744\_07 are located at opposite regions of the morphology space. Both of these galaxies exhibit RPS and PSB features (with A2744\_01 being primarily RPS and A2744\_07 primarily PSB) and are likely to be in an intermediate phase between the two types. Their extreme locations in morphology space therefore suggests that the visual indicators of the disturbance are at their highest toward the end of the RPS phase, and are quick to vanish after the stripping ceases.

\subsection{Centroid Variance Method}

Each of the above methods provide useful ways to quantify various aspects of the morphologies, however the disturbed morphologies of the ram-pressure stripped sample do not generally occupy specific regions of any given parameter space, but overlap with the general cluster population.

Here we test methods to detect the low surface brightness tails and disturbances characteristic of ram-pressure stripping interactions, with the aim of developing a criterion which is tailored to be more sensitive to such objects. To this end, we experiment with measuring the variation between the centroids of flux isocontours to quantify inconsistencies in the light distribution of the galaxy. RPS is often characterized by offset low surface brightness emission, which will cause the centroids of isocontours taken at different flux levels to vary more than they would in an undisturbed galaxy.

The technique is summarized as follows. We first mask the galaxy with the segmentation map, and take the range of flux values between the minimum and maximum values in the masked image. We ignore the lowest 20\% of this range, as the faintest contours tend to be clipped by the segmentation mask and do not reflect the shape of the galaxy's light distribution. We draw a set of 8 contours, between 20\%-99\% of the flux range, and for comparison an additional set of 8 contours focused only on the outer regions of the galaxy between 20\%-50\% of the flux range.

For each flux value we draw an isocontour of the emission, and obtain the position of the non-flux-weighted centroid of that contour using the {centre-of-mass} method from the python package \textsc{scipy}. For axisymmetric emission, such as an undisturbed circular or elliptical gaussian, the centroids of every contour would be expected to lie in exactly the same location. Any disturbances or asymmetries in the light distribution will manifest as variations in the centroid locations.

We then calculate the variance and covariance of the set of coordinates over all of the centroids, normalised by the galaxy's half-light radius, to quantify the movement of the centroids across the different flux thresholds, using the equations:

\begin{align*}
\mathrm{centroid\ variance} &= \frac{\sigma^{2}(\mathrm{X}) + \sigma^{2}(\mathrm{Y})}{\mathrm{r}_\mathrm{e}}\\ 
\mathrm{centroid\  covariance} &= \frac{|\mathrm{cov}(\mathrm{X},\mathrm{Y})|}{\mathrm{r}_\mathrm{e}}
\end{align*}

Where X and Y are the arrays of x and y coordinates of the flux isocontour centroids and $\mathrm{r}_\mathrm{e}$ is the galaxy half light radius

A higher variance indicates that the distribution of flux is non-uniform or asymmetric, whilst a high covariance indicates that the disturbance is along a particular direction. We found that the variance of the centroid offers a promising indicator of disturbance. The covariance performs similarly well, with slightly less separation between disturbed an undisturbed morphologies.

\begin{figure}
    \centering
    \includegraphics[width=0.5\textwidth]{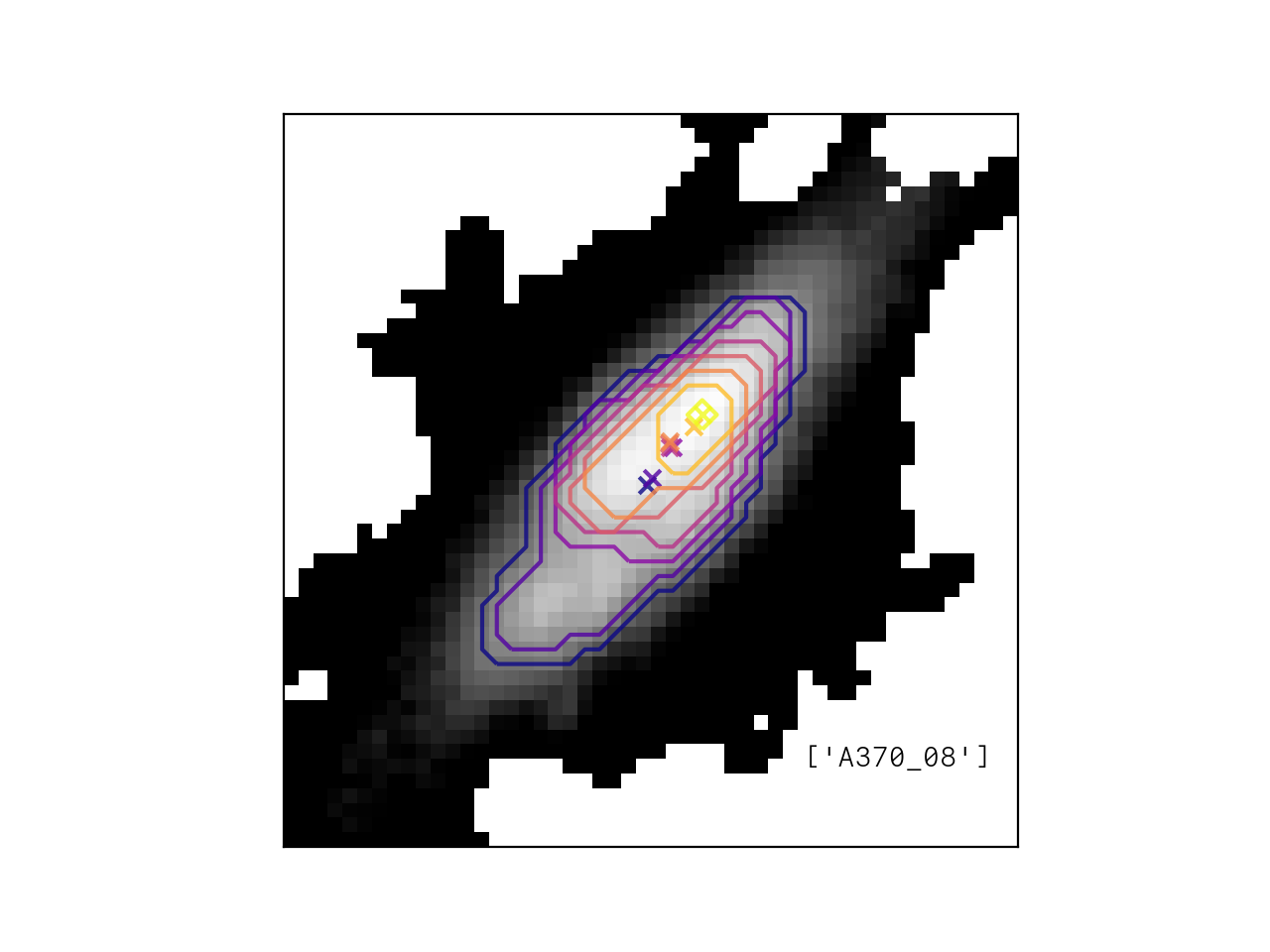}
    \caption{Example of centroid variance measurement on the RPS galaxy A370\_08. The F606W image of the galaxy is shown overlaid with the flux isocontours and their corresponding centroids as x markers.}
    \label{fig:contours}
\end{figure}

An example is shown in Figure~\ref{fig:contours} for A370\_08, which shows the contours between 20\% and 99\% of the range of flux values, and the centroids of each contour as x markers colored accordingly. The skewed light distribution resulting from the disturbance pushes the centroids of the lower surface brightness emission towards the lower left of the figure relative to the central peak, resulting in an increased variance.

A scatter plot of the centroid variance in the outer region vs the full range is shown in the lower left panel of Figure~\ref{fig:morphology}. The figure shows that the majority of galaxies in the RPS sample have significantly higher centroid variances than the rest of the cluster sample, using either range of flux values to define the contour levels. The sample of PSB galaxies do not appear to be distinct from the rest of the cluster galaxies, as is the case for the other morphology parameters. The red cluster galaxies are fairly concentrated with low centroid variances, with the blue galaxies and galaxies with F606W-F814W$<0.6$ having slightly higher centroid variances in the outer regions in comparison. The distinction between the stripped galaxies and the rest of the cluster sample suggests that the variance in the non flux-weighted centroid of emission across different flux thresholds is a promising indicator of disturbed morphologies and could feasibly be used to detect galaxies of interest. It is possible that disturbances due to non-RPS processes, e.g. tidal interaction, could affect the centroids in a similar way, however on inspection of the non-RPS objects which are close to our RPS sample in the figure, we do not see evidence of gravitational disturbances, and a dedicated sample would be required to test the classification of gravitationally interacting galaxies using this method.
The plot of the covariances can be found in Figure~\ref{fig:c_c} in Appendix~\ref{sec:appendix}, for reference.

To test the success of the centroid variance parameters at resolving the RPS galaxies from the cluster populations, we carried out 2 sample K-S tests for each morphology parameter.

We found that at the 5\% significance level, the \textit{concentration} ($p=0.12$) and \textit{gini} ($p=0.16$) parameters cannot distinguish the RPS galaxies from the distribution of undisturbed cluster galaxies, whereas for the \textit{asymmetry} ($p=2.84\times10^{-5}$), \textit{M$_{20}$} ($p=0.04$), outer centroid variance ($p=4.13\times10^{-6}$) and full centroid variance ($p=7.11\times10^{-7}$) parameters, the RPS galaxies are distinct from the undisturbed cluster population.

To test the selection of RPS galaxies using this diagnostic, we draw a line approximately separating the distinct clump of galaxies in the top right from the rest of the sample, described by the equation:

\begin{align*}
\log_{10}&(\mathrm{outer\ CV}) > -2\times\log_{10}(\left(\mathrm{full\ CV}\right))-2.2
\end{align*}

Where outer CV and full CV refer to the outer centroid variance and full centroid variance respectively. This criterion yields a sample of 10 galaxies, of which 4 are long-tailed RPS galaxies, 3 are short-tailed RPS galaxies, and 3 are non-RPS cluster galaxies.

We calculated the precision and recall of this selection criterion, defined as:

\begin{align*}
\mathrm{precision}&=\frac{\mathrm{true\ positives}}{\mathrm{true\ positives} + \mathrm{false\ positives}}\\
\mathrm{recall}&=\frac{\mathrm{true\ positives}}{\mathrm{true\ positives} + \mathrm{false\ negatives}}
\end{align*}

finding a precision of 0.70 and a recall of 0.58.


\subsection{Principal Component Analysis}\label{sec:PCA}

We applied principal component analysis (hereafter PCA) in order to visualise the distribution of these galaxies and investigate whether similar galaxies are grouped together in the combined morphology space. PCA reduces multidimensional parameter space by finding the ``principal components" of covariant parameters, in order to explore relations between different parameters, or simplify the visualization of a large number of parameters which may all scale according to some common underlying property. In this case, each of the morphology parameters gives us complementary, but overlapping information about the shape of a galaxy. By normalizing each of the different quantities to negate the effects of scale, and transforming them into a reduced parameter space, we can retain the maximum amount of information with a simplified set of quantities.
PCA transforms a data into a set of orthogonal eigenvectors, or principal components, which are linear combinations of the input parameters. These are arranged such that the maximum amount of variance from the original data is contained in the first few eigenvectors of the transformed dataset.

To select the number of relevant principal components, we use the rule proposed by \citet{Kaiser1960}, whereby components are rejected if they contain less than the expected variance of uncorrelated variables. In this case, with 9 variables at play, each would contain 11.1\% of the total sample variance if no correlation was present. We find that 3 principal components are above this threshold, which altogether contain 67\% of the total sample variance (PC1: 37\%, PC2: 16\%, PC3: 13\%). Notably, the first principal component contains more than double the sample variance of each of the other components. The variances of the top 5 principal components yielded by the PCA are shown graphically in Figure~\ref{fig:pca_variance} in Appendix~\ref{sec:appendix}.

The principal components resulting from the PCA are described in Table~\ref{tab:PCA}. The input parameters (each of the morphology quantities) are given in the first column of the table. To obtain each of the principal components PC1, PC2 and PC3, the morphology parameters for a given galaxy are scaled by their corresponding weights (given in columns PC1, PC2 and PC3 of table~\ref{tab:PCA} respectively) and combined in summation (i.e.: $\mathrm{PC1}=-0.276\times\textit{concentration} + 0.16\times\textit{asymmetry} - 0.004\times\textit{gini} + ...$). The first principal component, PC1, which contains significantly more of the sample variance than the other components, is most influenced by the outer centroid variance and full centroid variance parameters, with weights of 0.474 and 0.457 respectively. PC2 is driven mostly by asymmetry and gini, whilst PC3 is influenced mostly by the full centroid covariance and the concentration. All of the components, however, are non-negligibly influenced by several other parameters in addition to these.

\begin{figure}
    \centering
    \includegraphics[width=0.5\textwidth]{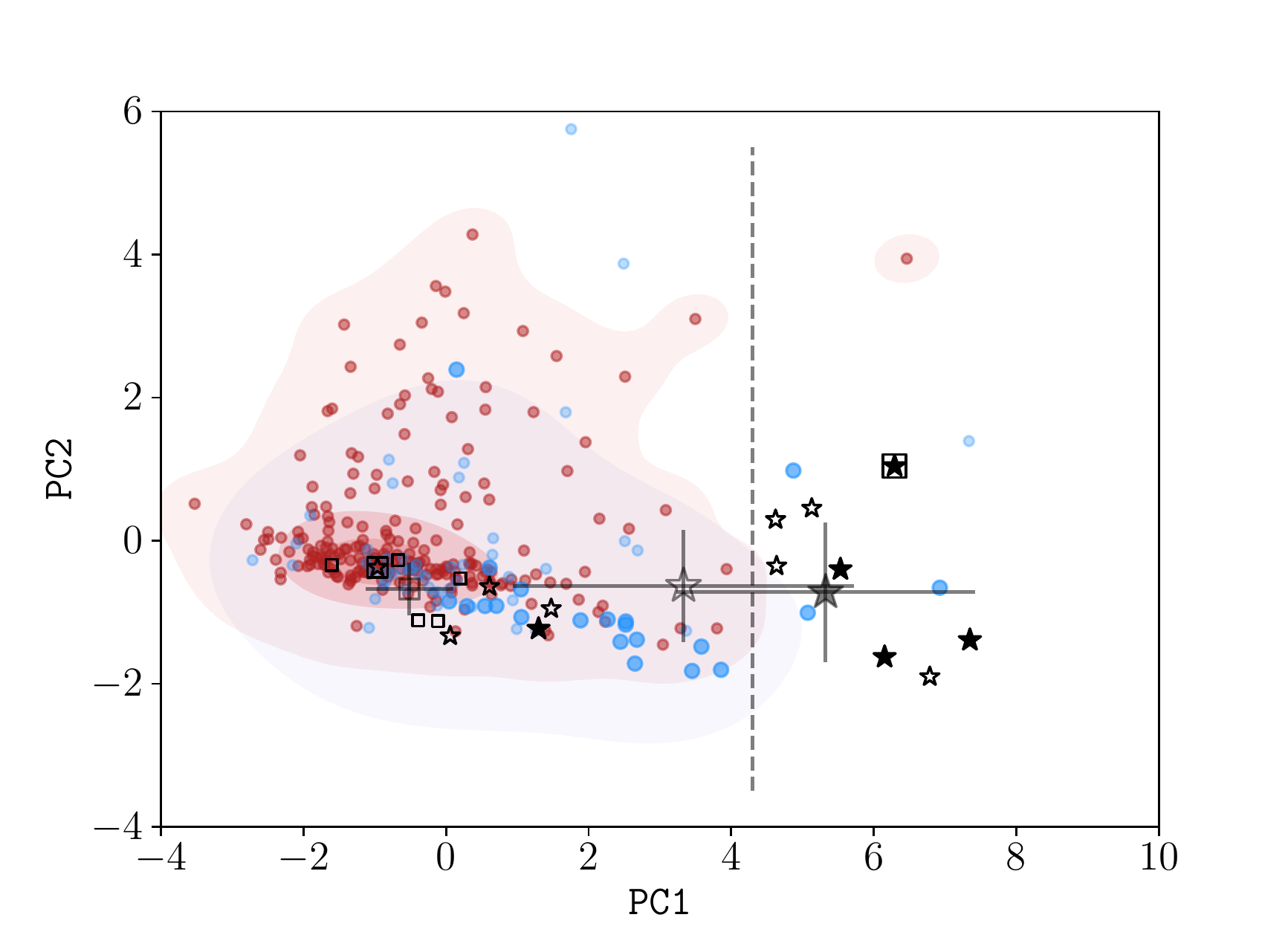}
    \includegraphics[width=0.5\textwidth]{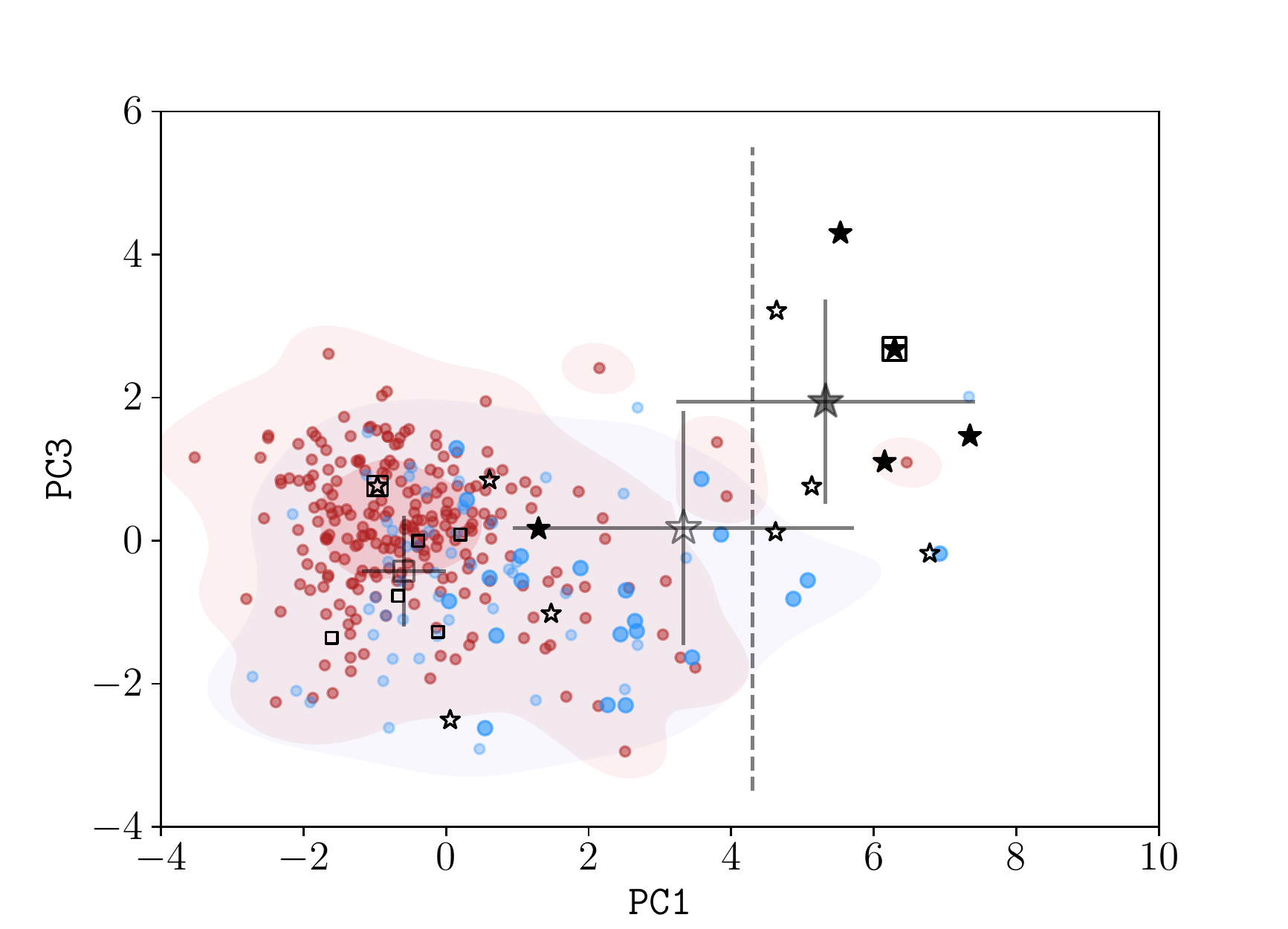}
    \caption{Scatter plots of the three components resulting from the principal component analysis described in the text, for all cluster members in A2744 and A370 combined. Data points are colored or indicated by symbols as described in Figure~\ref{fig:morphology}. The median and the standard deviation of the two components within each category are shown as error bars marked with the relevant color or symbol. The vertical dashed line indicates the $\mathrm{PC1}>4.3$ threshold suggested here to select potential ram-pressure stripped galaxies.}
    \label{fig:pca}
\end{figure}

The sample of galaxies is displayed reprojected into the resulting principal component space in Figure~\ref{fig:pca}. These 3 components on their own do not describe physical properties, but help to visualize any groupings of galaxies in higher dimensional parameter space in a simplified plot.

\begin{table}
    \centering
    \begin{tabular}{l|r
    r r r}
        &\multicolumn{3}{c}{Weights $\omega_i$}\\
        Input Variable $x_i$ & PC1 & PC2 & PC3\\\hline
        concentration & -0.276 & -0.073 & 0.481\\
        asymmetry & 0.160 & 0.654 & -0.126\\
        gini & -0.004 & 0.589 & 0.387\\
        M$_{20}$ & 0.282 & 0.370 & -0.302\\
        $\mathrm{Log}_{10}$(centroid var.) (outer) & 0.474 & -0.119 & -0.039\\
        $\mathrm{Log}_{10}$(centroid covar.) (outer) & 0.436 & -0.113 & -0.011\\
        $\mathrm{Log}_{10}$(centroid var.) (full) & 0.457 & -0.096 & 0.324\\
        $\mathrm{Log}_{10}$(centroid covar.) (full) & 0.315 & -0.038 & 0.593\\
        F606W-F814W color & -0.309 & 0.211 & 0.234\\
    \end{tabular}
    \caption{Table of weights $\omega_i$ assigned to each of the input variables $x_i$ to yield the principal components in the linear combination $\omega_1x_1 + \omega_2x_2 + \omega_3x_3 + ... $}
    \label{tab:PCA}
\end{table}

The three components are shown in Figure~\ref{fig:pca}, with PC1 compared with PC2 in the upper panel, and with PC3 in the lower panel.

It is clear from the figure that PC1 provides the strongest separation between the RPS sample and the cluster galaxies, with the vertical line at PC1 = 4.3 drawn on the figure to mark our suggested threshold.

This criterion of $\mathrm{PC1} > 4.3$ selects 4/5 long-tailed RPS galaxies, 4/7 short tailed RPS galaxies, as well as 4 blue galaxies and 1 red galaxy. We calculated a precision of 0.62 and a recall of 0.67 for the threshold of 4.3, and additionally show the two diagnostics calculated over a range of threshold values shown in Figure~\ref{fig:prec_rec} in Appendix~\ref{sec:appendix}. In comparison, the precision of this selection is slightly lower than when selecting galaxies using only the centroid variance measurements, but the recall is higher, i.e., the PCA selection retrieved more of the known RPS sample.

We further inspected the non-RPS sample objects and found that two of the blue objects in this region of the diagram are clumps which are close in location and velocity to A2744\_01 and A370\_08, strongly suggesting that they are clumps of material associated with those galaxies, which have been flagged as distinct objects by the source extraction but indeed are considered part of the jellyfish tail by our MUSE analysis \citep{Moretti2022}. The latter of those two objects, the clump to the south of A370\_08, is likely to be the object identified as CL49 in \citet{Lagattuta2017,Lagattuta2019} which the authors also concluded was a clump of material detached from the main galaxy. Two sources, the red object and one of the blue objects, were found to be galaxies overlapping in projection. The last blue object was inspected and found to have associated emission lines in the MUSE image, however the object is very small and faint, and whilst it appears to disturbed in the HST image, the nature of the disturbance cannot be verified.




\begin{figure*}
    \centering
    \includegraphics[width=\textwidth]{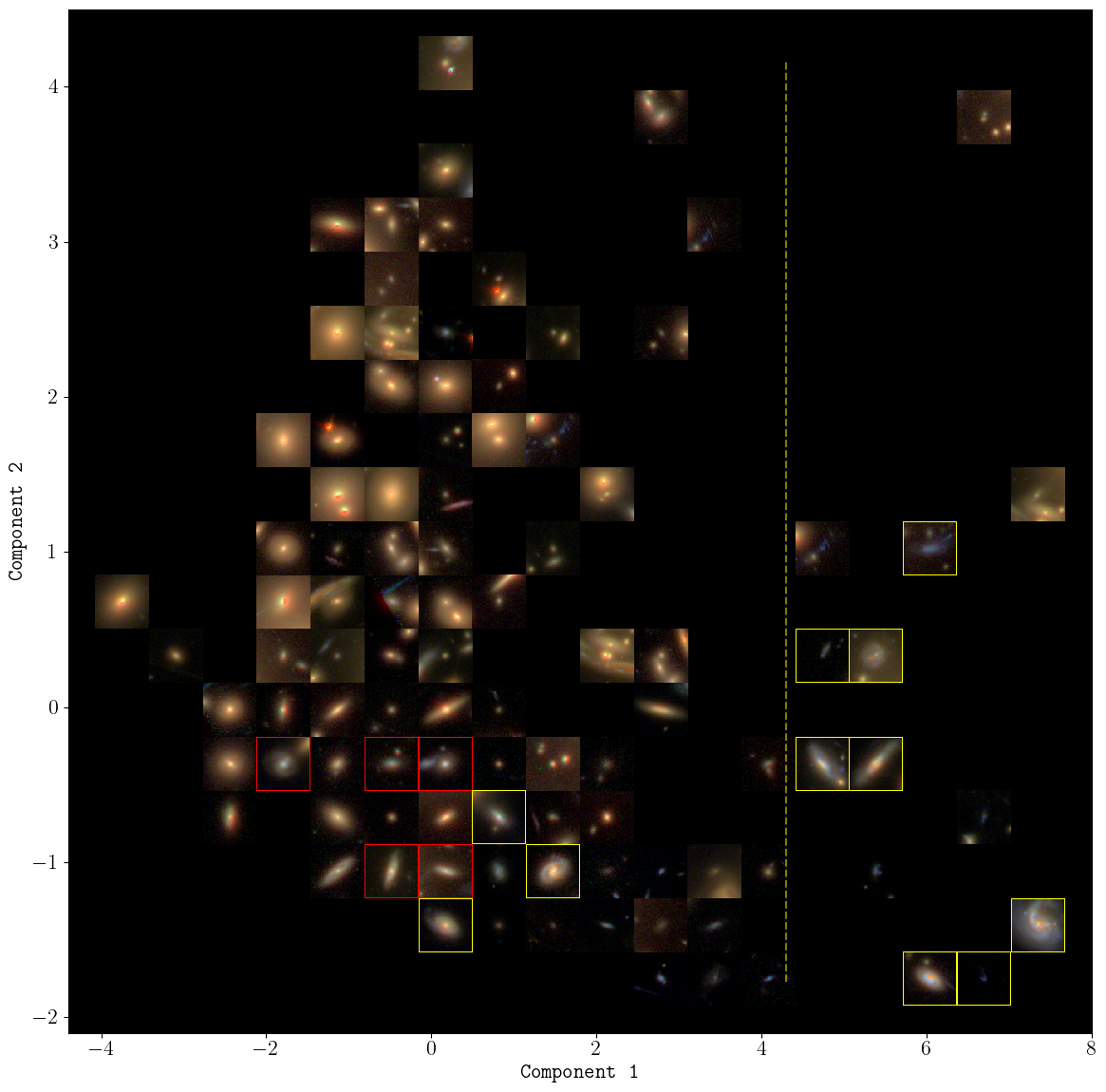}
    \caption{Grid of sample galaxies in the dimensionally reduced space yielded by the principal component analysis described in section~\ref{sec:morphology}. At each location on the grid, an example galaxy is shown in order to visualize the different morphologies which occupy different regions of the reduced morphology space. RPS and PSB galaxies are marked with yellow and red borders respectively. If multiple types occupy a bin, priority is given to showing RPS galaxies, then PSB galaxies, then undisturbed cluster galaxies. The yellow dashed line indicates the $\mathrm{PC1} > 4.3$ threshold described in the text, above which the majority of RPS galaxies are found.}
    \label{fig:pca_grid}
\end{figure*}

The distribution of galaxies across PC1 and PC2 is also shown in Figure~\ref{fig:pca_grid}, with the space divided into bins and an example galaxy shown for each bin to indicate typical morphologies corresponding to that combination of parameters. In general, the lower left corner of the figure appears to contain the majority of undisturbed spiral galaxies as well as elliptical and spheroidal galaxies. Moving upwards toward the top left corner of the figure, many of the objects appear to be in increasingly crowded environments or have close companions, which may be impacting their morphology parameters. Moving towards the bottom right, the galaxies appear to have increasingly disturbed morphologies, which is where most of our RPS sample is located.

\section{Discussion and Summary}\label{sec:discussion}

We have analysed a sample of 12 RPS galaxies and 6 PSB galaxies within two clusters at intermediate redshift, A2744 and A370. We have compared several characteristics of the RPS and PSB galaxies with the general cluster population, specifically, their orbital information, their distribution within cluster substructures and environment, and their morphologies.

We found that the general cluster population in the observed field of A2744 follows a bimodal distribution, with the two components having similar velocities to the regions described as the SMRC and the NC in \citet{Owers2011}. All of the RPS galaxies, and all but one of the PSB galaxies are located in the blueshifted structure, along with a significantly higher fraction of blue galaxies in comparison with the redshifted component. Together, this is indicative that the collision of the galaxies in the CTD with the X-ray gas associated with the SMRC is responsible for the stripping being experienced by the RPS galaxies. The difference in blue fractions between the substructures may be due to an excess of blue galaxies in the blueshifted component caused by increased star formation due to weak stripping, or a dearth of blue galaxies in the redshifted component resulting from quenching during a previous merger event.

We find that in A370, the RPS galaxies and PSB galaxies are more evenly distributed in phase-space and the majority are not residing in any substructures. Whilst A370 is also a merging cluster, we do not see any evidence that the observed RPS galaxies are the result of the merger, rather that they are more likely to be isolated infalling galaxies.

We analysed the ICM X-ray emission at the locations of the cluster galaxies, and found that in both clusters the PSB galaxies reside in regions of higher ICM X-ray flux compared with the blue cluster members and RPS galaxies. In A370 the RPS galaxies are also located at higher ICM X-ray fluxes than the blue cluster galaxies, but not as high as the PSB galaxies.
The location of the PSB galaxies in regions of higher ICM X-ray flux, which scales with gas density, is consistent with the population of PSB galaxies being produced, at least in part, by ram-pressure interactions.

Finally, we implemented several measures to quantify the morphologies of the galaxies and compared the results for the different samples. We utilised \textit{concentration}, \textit{asymmetry}, \textit{gini} and \textit{M20}, and also tested whether the variance and covariance of the emission centroid of different flux isocontours could be a useful measure of the disturbance caused by RPS.

We found that the most effective standalone measure of the morphology was the centroid variance, which shows promise as a potential measure to detect disturbed morphologies. By combining the different parameters using principal component analysis, the scatter was further reduced and the separation of both weakly and strongly disturbed galaxies from the general cluster population was made clearer than when using any of the individual morphological quantities alone. This could have practical applications for filtering through large broad-band imaging datasets for potentially disturbed galaxies prior to manual inspection to determine the cause of the disturbance, reducing the workload on human classifiers. These kinds of automated techniques will open up the possibility of detecting samples of RPS candidates from huge catalogs of survey data, including, as we have found, cases where the disturbance is minimal in the broad-band images. By expanding the known RPS sample and obtaining snapshots of galaxies from first infall to fully quenched, we will be able to further explore the process of ram-pressure stripping and its impact on galaxies in clusters.


\section*{Acknowledgements}

We wish to thank D. J. Lagattuta for valued comments and suggestions. We are grateful to the anonymous referee who helped us clarify and strengthen this analysis.
This project has received funding from the European Research Council (ERC) under the European Union’s Horizon 2020 research and innovation program (grant agreement No. 833824, GASP project). We acknowledge funding from the INAF main-stream funding programme (PI B. Vulcani) and the Italian PRIN-Miur 2017 (PI A. Cimatti). YJ gratefully acknowledges support from the ANID BASAL project FB210003. \software{Astropy \citep{astropy:2013, astropy:2018}, statmorph \citep{Rodriguez-Gomez2019}, \textsc{sinopsis} \citep{Fritz2017}, \textsc{highelf}}




\bibliographystyle{aasjournal}
\bibliography{hiz} 



\appendix
\section{Additional Figures}\label{sec:appendix}
\begin{figure*}[h]
    \centering
    \includegraphics[width=0.9\textwidth]{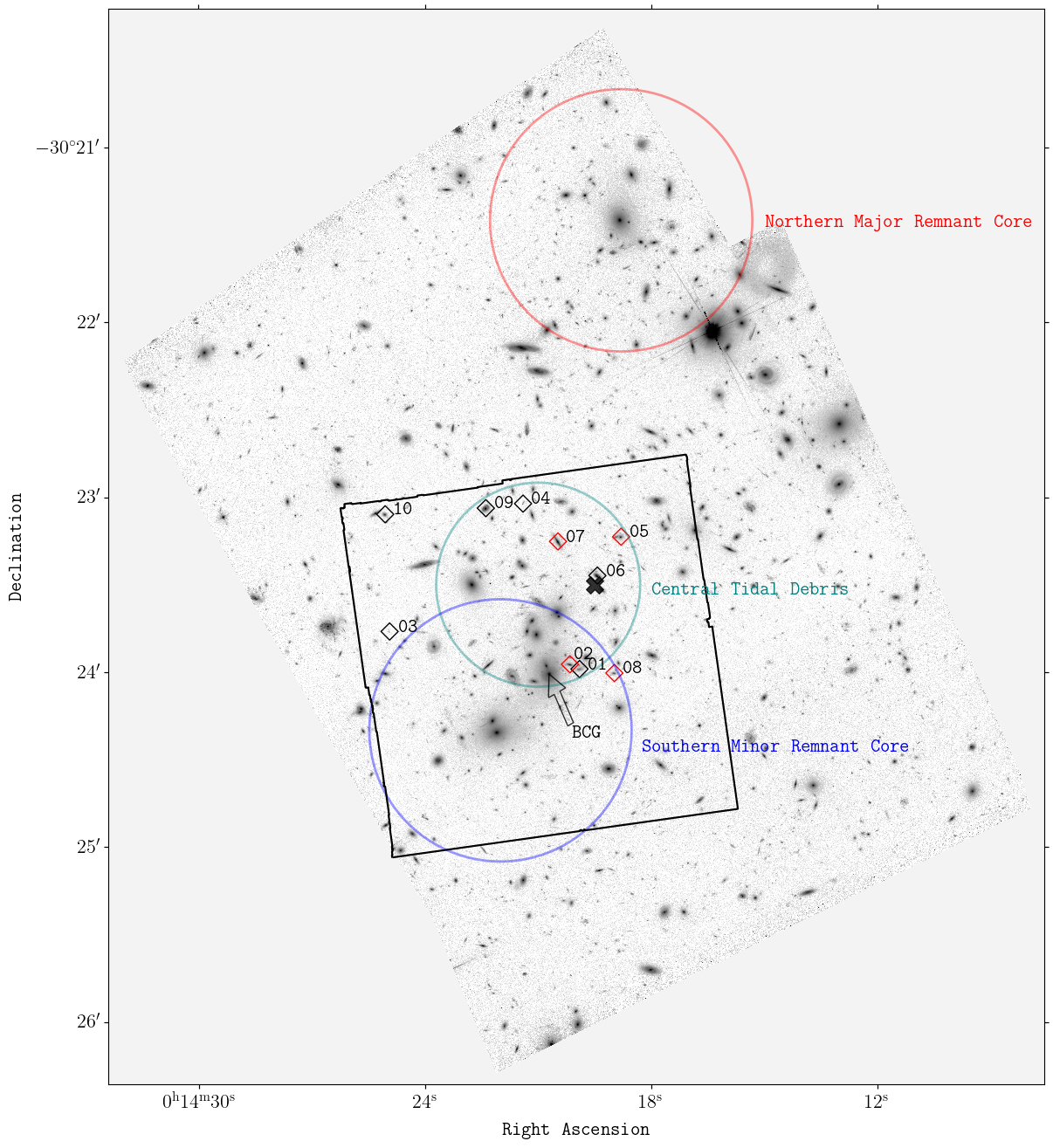}
    \caption{Overview of A2744 showing the HST F606W image of the cluster, with the MUSE footprint shown by the black square. Locations of the different cluster regions as defined in \citet{Owers2011} are overlaid as colored circles, the RPS and PSB galaxies are shown by the black and red diamonds, respectively. The arrow marks the BCG used as the center for this analysis and the black X marks the X-ray surface brightness peak measured by \citet{Owers2011}}
    \label{fig:A2744_footprint}
\end{figure*}

\begin{figure*}[h]
    \centering
    \includegraphics[width=0.9\textwidth]{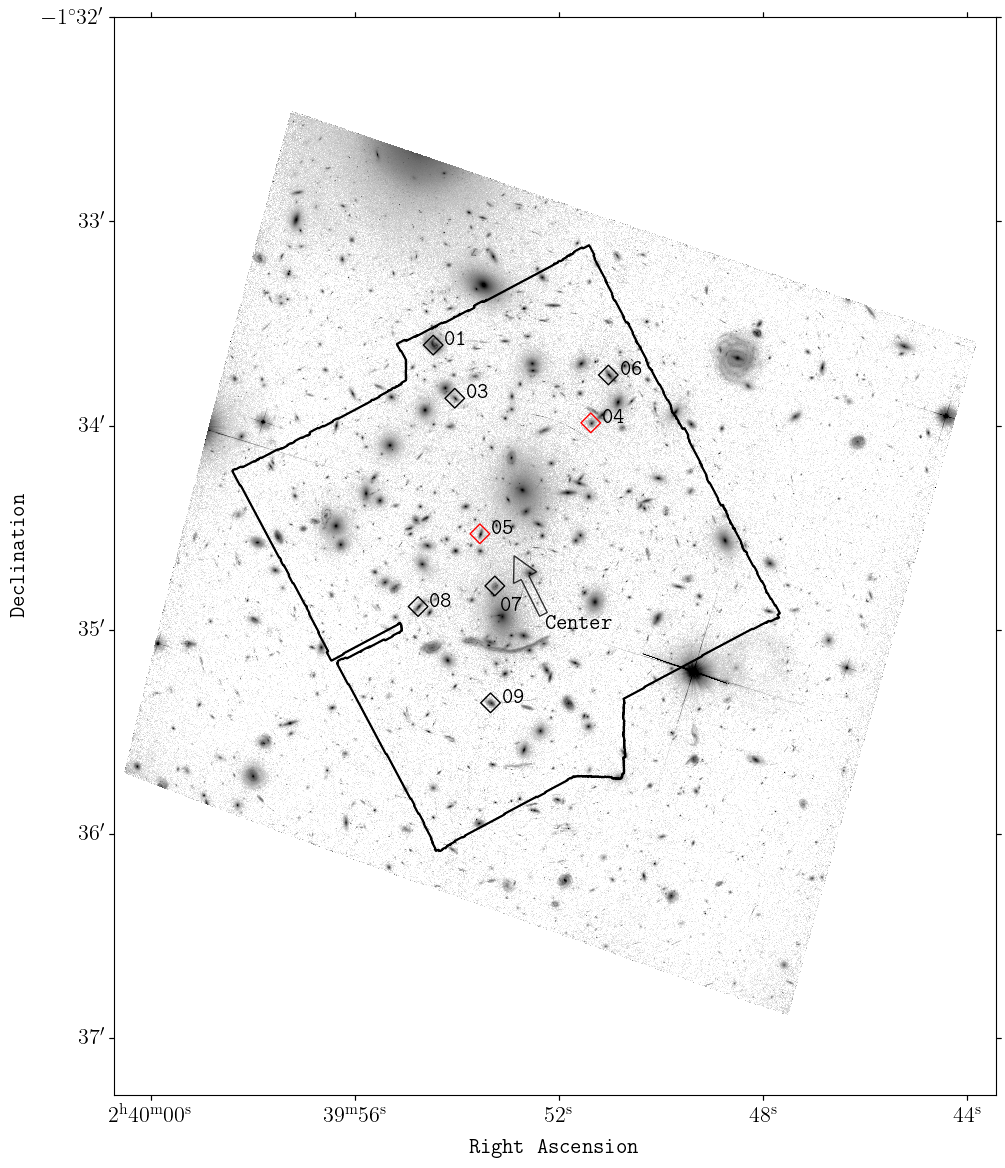}
    \caption{Overview of A370 showing the HST F606W image of the cluster, with the MUSE footprint shown by the black square. RPS and PSB galaxies are shown by the black and red diamonds, respectively. The arrow marks the point between the two BCGs used as the center for this analysis.}
    \label{fig:A370_footprint}
\end{figure*}

\begin{figure}[h]
    \centering
    \includegraphics[width=0.5\textwidth]{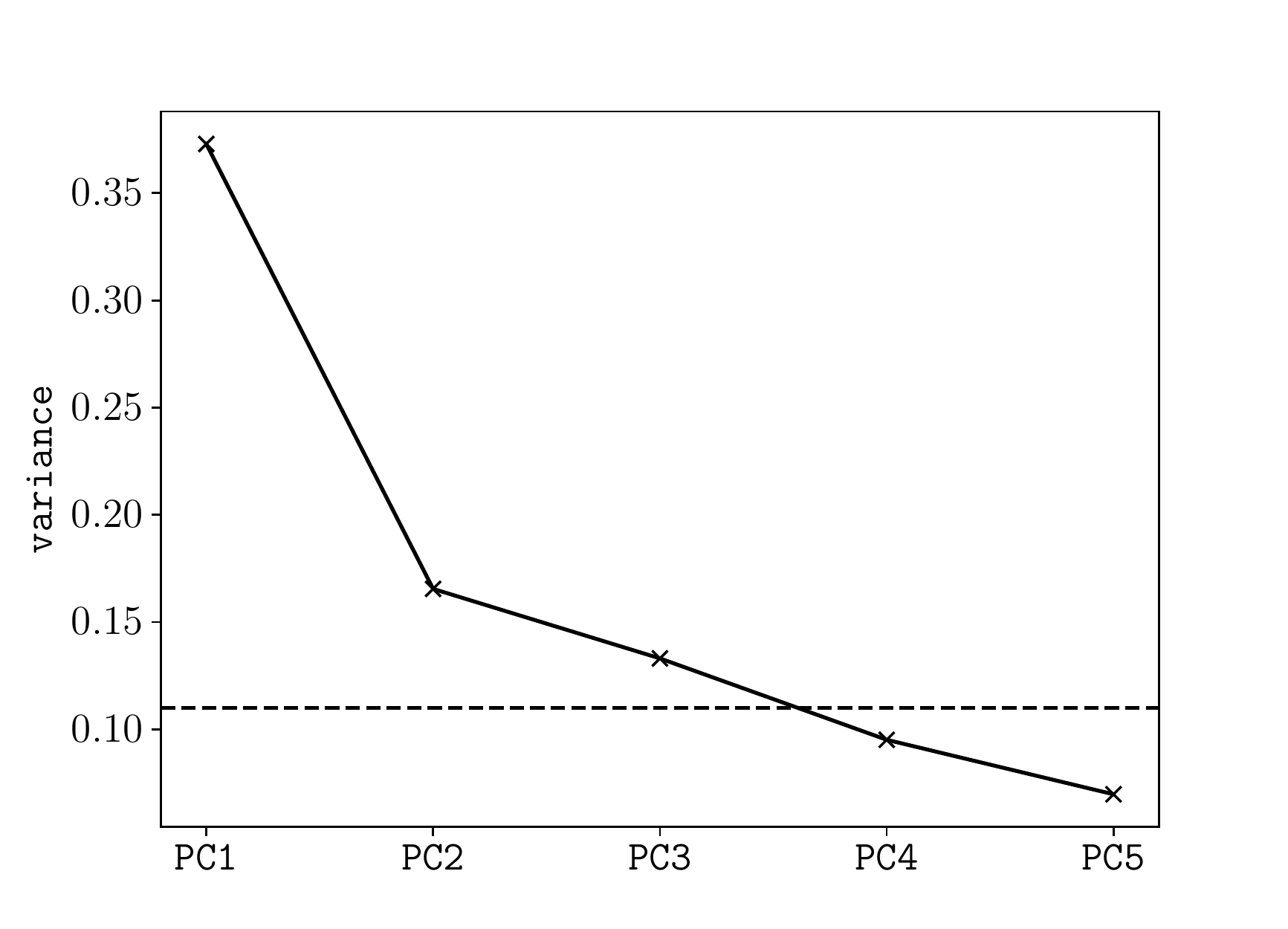}
    \caption{Variance of the top 5 principal components yielded by the PCA. The horizontal dashed line shows the cutoff of 11.1\%, below which a principal component was not used, as described in Section~\ref{sec:PCA}.}
    \label{fig:pca_variance}
\end{figure}

\begin{figure}[h]
    \centering
    \includegraphics[width=0.5\textwidth]{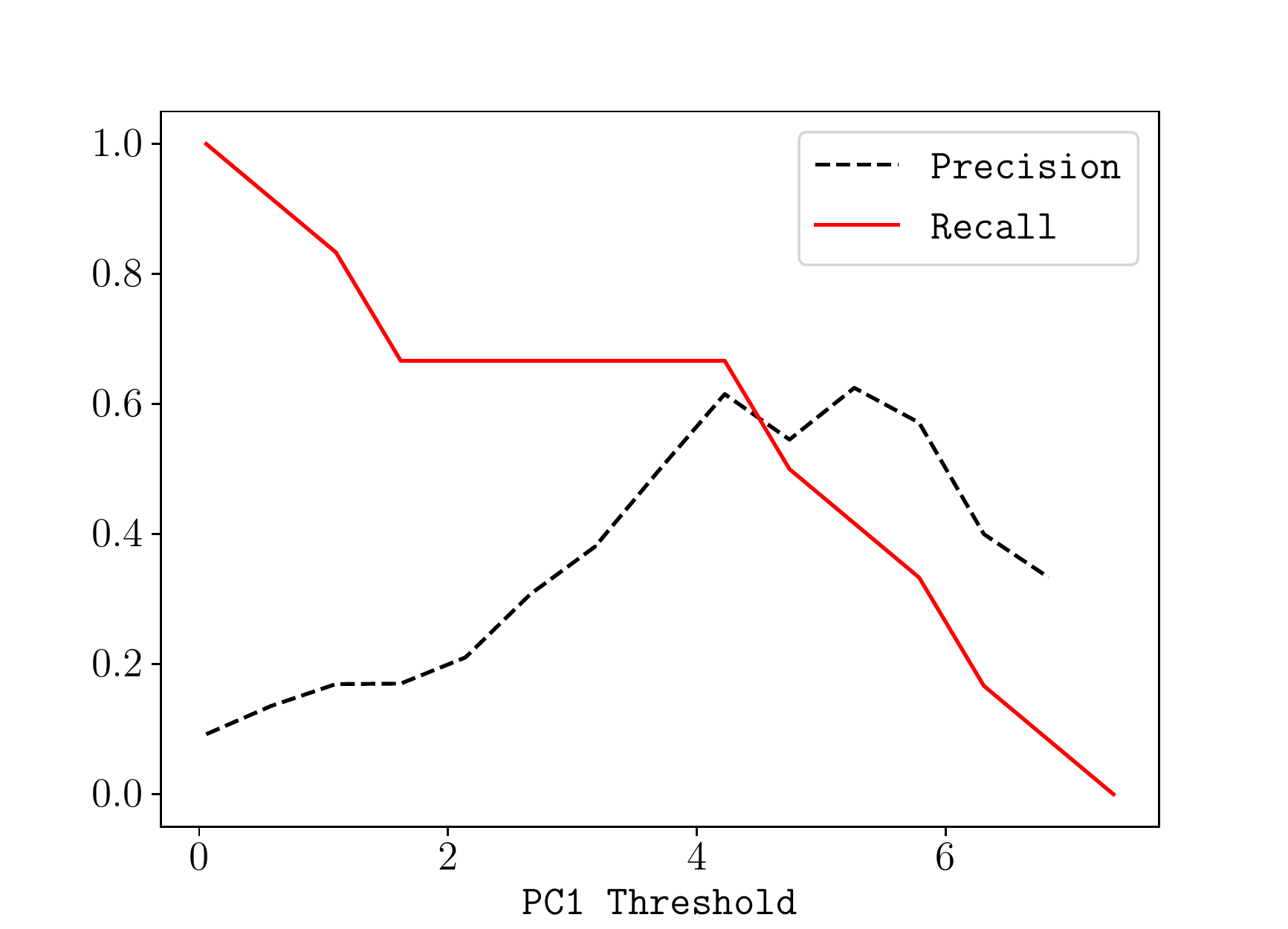}
    \caption{Values of precision and recall calculated across the range of selection thresholds for PC1. The selected value of 4.3, chosen in the analysis, is close to the crossover between the two diagnostics as well as the peak in the precision.}
    \label{fig:prec_rec}
\end{figure}

\begin{figure}[h]
    \centering
    \includegraphics[width=0.5\textwidth]{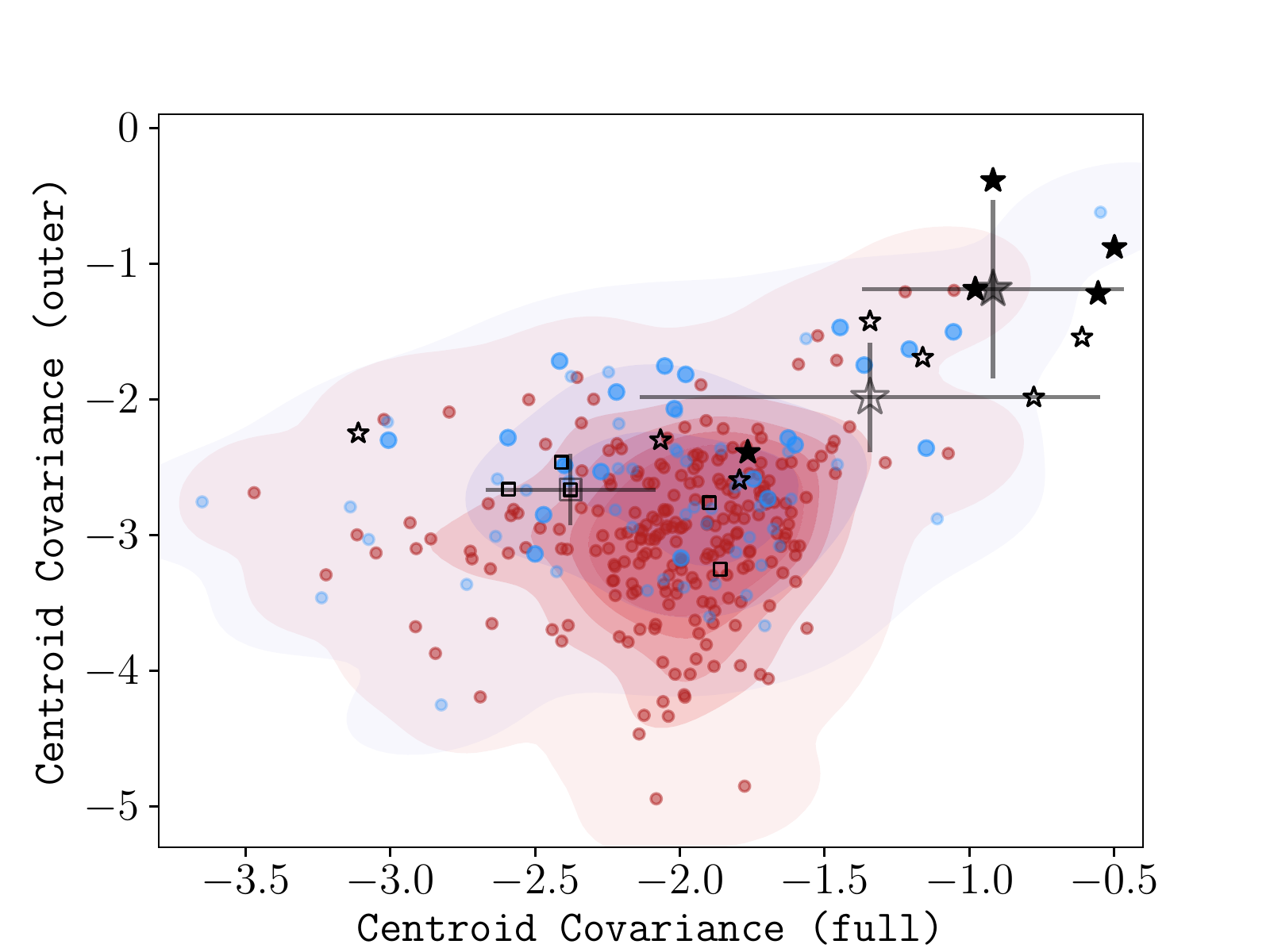}
    \caption{Plot of the outer and full centroid covariance parameters for galaxies within A2744 and A370 combined, derived as described in the text. The median and the standard deviation of the centroid covariance within each category are shown as error bars marked with the relevant color or symbol. Galaxies are coloured accordingly as described in the colour magnitude diagrams, Figure~\ref{fig:CM_both}, and marked as described in Figure~\ref{fig:morphology}.}
    \label{fig:c_c}
\end{figure}

\begin{figure}[h]
    \centering
    \includegraphics[width=\textwidth]{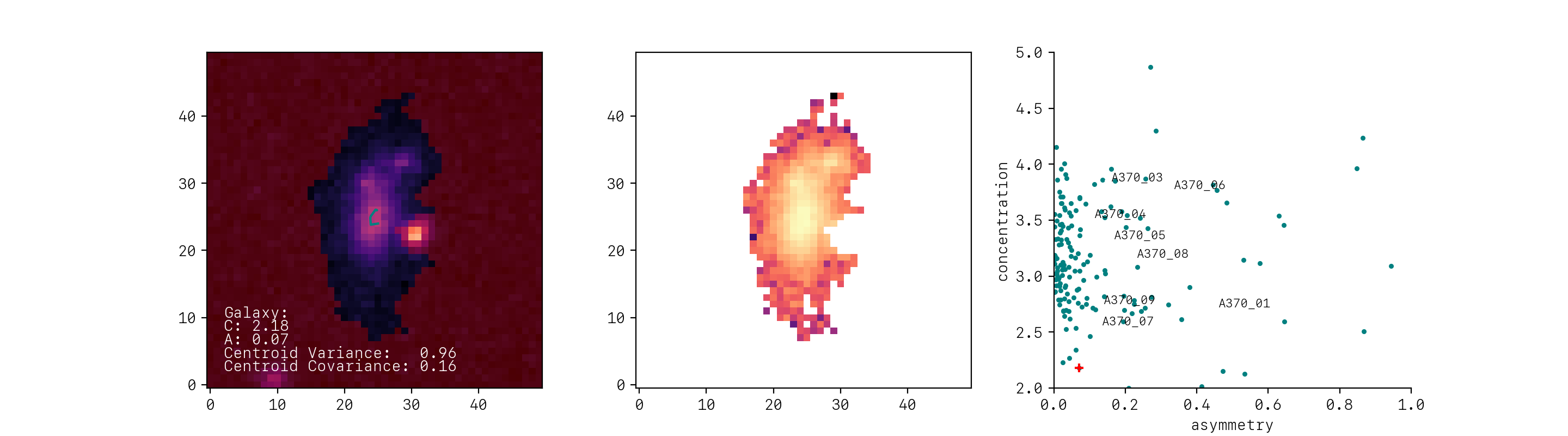}
    \includegraphics[width=\textwidth]{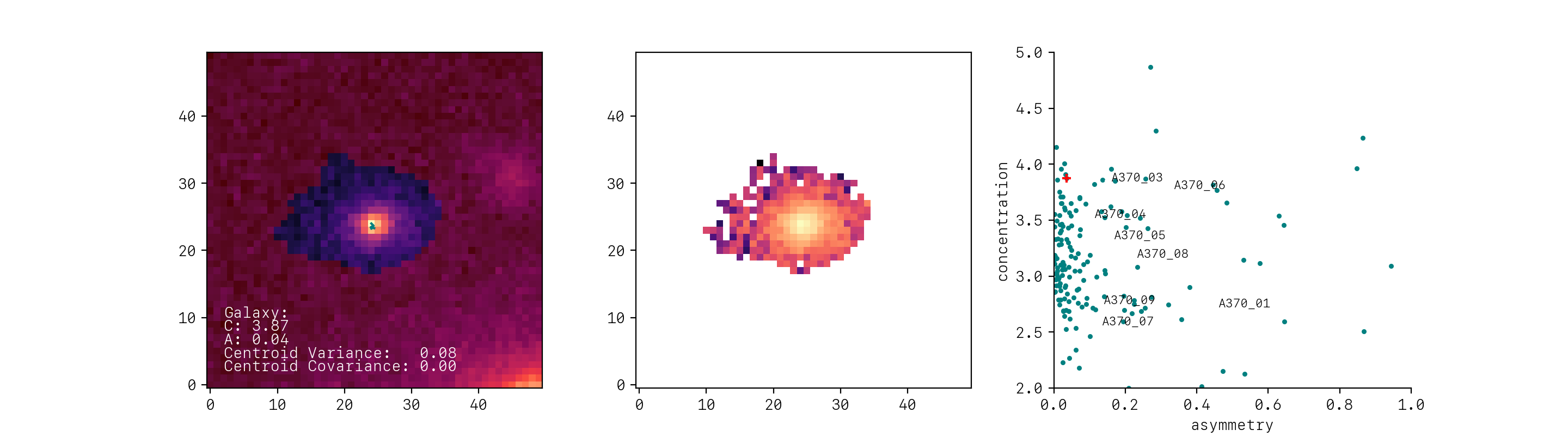}
    \includegraphics[width=\textwidth]{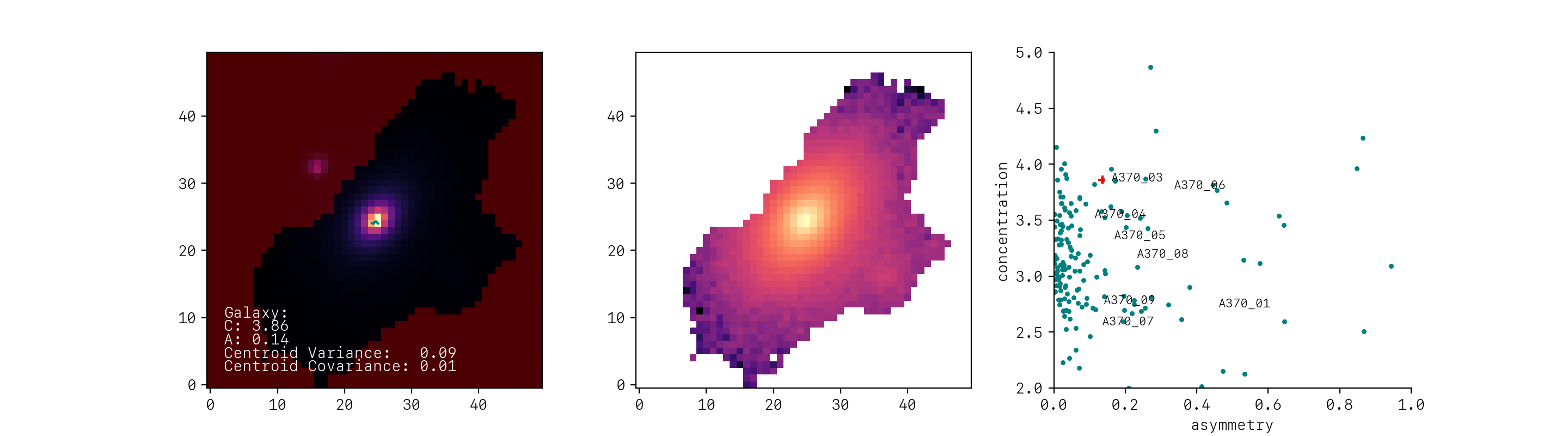}
    \includegraphics[width=\textwidth]{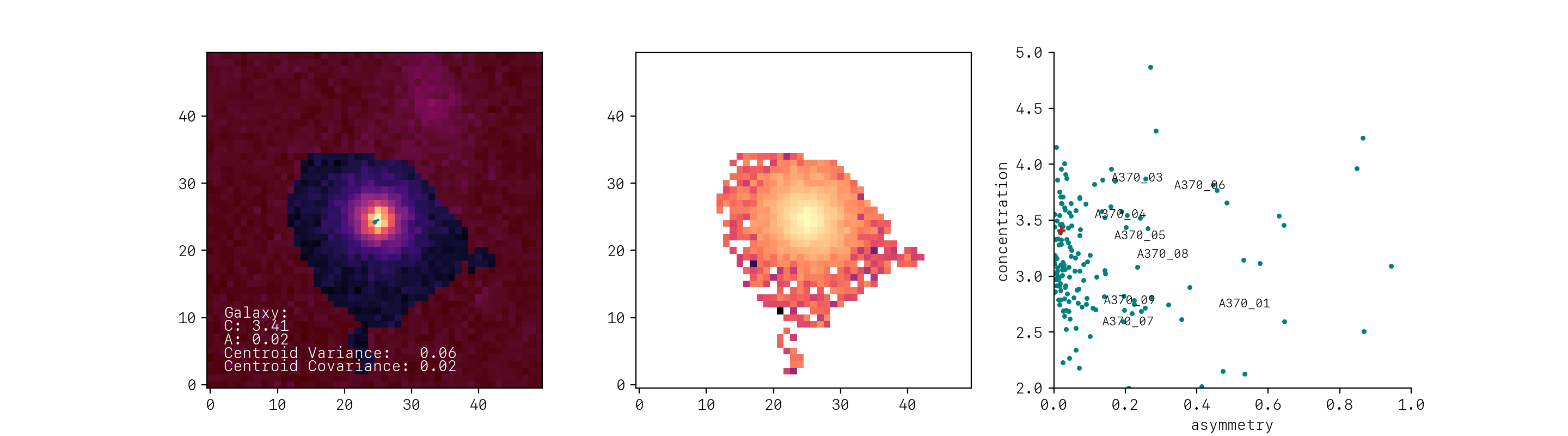}
    \caption{Example morphology diagnostics and segmentation maps, showing some cases where a galaxy has a close neighbor and some deblending has been required to produce the segmentation map.}
    \label{fig:segmentation}
\end{figure}


\label{lastpage}
\end{document}